\documentclass[11pt]{article}
\usepackage{jheppub}

\usepackage{siunitx}
\usepackage{jheppub}
\usepackage{array}
\usepackage{float}
\usepackage{booktabs}
\usepackage{makecell}
\usepackage{caption}
\usepackage{amsmath}
\usepackage{graphicx,tabularx}
\usepackage{gensymb}
\usepackage{url}
\usepackage{footmisc}
\usepackage{amsfonts}
\usepackage{cancel}
\usepackage{color}
\usepackage{multirow} 
\usepackage{amssymb}
\usepackage{pifont}
\usepackage{epstopdf}
\usepackage{slashed}
\usepackage{comment}
\usepackage{booktabs} 
\usepackage{subfigure}
\usepackage{mathrsfs}
\usepackage{subcaption}
\usepackage[toc,page]{appendix}
\usepackage{mathtools}
\usepackage{romannum}
\usepackage[normalem]{ulem}
\usepackage{bbold}
\usepackage{xcolor}
\usepackage{tikz-feynman}
\usepackage{tikz}
\usepackage{placeins}
\usepackage{hyperref}
\usepackage{etoolbox,orcidlink}
\usepackage{feynmp-auto}
\usepackage{xcolor}
\usepackage[capitalise]{cleveref}

\allowdisplaybreaks

\usepackage{accents}

\def\ts{\tau^+\tau^-}
\def\mm{\mu^+\mu^-} 
\def\gg{\gamma\gamma}
\def\cc{c\bar c}

\def\beq{\begin{equation}}
\def\eeq{\end{equation}}
\title{Neutrino Fluxes at a Muon Collider} 

\author[]{Francis M. Burk${}^1$\orcidlink{0009-0001-7398-8211},}
\affiliation[1]{Pittsburgh Particle Physics, Astrophysics, and Cosmology Center, Department of Physics and Astronomy, University of Pittsburgh, Pittsburgh, PA 15260\looseness=-1}
\emailAdd{frb25@pitt.edu}

\author[]{Tao Han${}^1$\orcidlink{0000-0002-5543-0716},}
\emailAdd{than@pitt.edu}

\author[]{Wolfgang Kilian${}^{2,1}$,}
\affiliation[2]{Theoretische Physik 1, Center for Particle Physics Siegen,
Universität Siegen, 57068 Siegen, Germany}
\emailAdd{kilian@physik.uni-siegen.de}

\author[]{Felix Kling${}^3$\orcidlink{0000-0002-3100-6144},}%
\affiliation[3]{Universit\"at Bonn, Regina-Pacis-Weg 3, D-53113 Bonn, Germany}%
\emailAdd{fkling@uni-bonn.de}

\author[]{Joachim Kopp${}^4$\orcidlink{0000-0003-0600-4996},}%
\affiliation[4]{Johannes Gutenberg-Universit\"at Mainz, 55128 Mainz, Germany}
\emailAdd{fkling@uni-bonn.de}

\author[]{Zahra Tabrizi${}^1$\orcidlink{0000-0002-0592-7425},}%
\emailAdd{z\_tabrizi@pitt.edu}

\preprint{
\begin{flushright}
PITT-PACC-2510 \\
SI-HEP-2026-15 \\
MITP-26-036
\end{flushright}
}

\abstract{
While muon colliders are primarily considered precision machines for new physics searches at the energy frontier, they are also intense sources of high-energy neutrinos. Large fluxes of electron and muon neutrinos are produced in the decay of beam muons, especially along the straight sections around the experiments. In this work, we re-evaluate these neutrino fluxes using the latest design of a \SI{10}{TeV} muon collider and find that fluxes around the interaction region are almost two orders of magnitude higher than earlier estimates. We also compute the fluxes of neutrinos produced in $\mu^+\mu^-$ collisions, electromagnetic showers induced by electrons from muon decay, and interactions of neutrinos close to the collider ring. We find that these additional neutrino sources are non-negligible and would lead to a sizable number of neutrino interactions of all flavors, including several tau neutrino events, in a ton-scale detector placed in the forward direction. We discuss the physics opportunities offered by muon collider neutrinos in the context of QCD and nuclear physics, electroweak precision measurements, and searches for new physics.
}

\begin{document}

\titlepage
\maketitle

\section{Introduction}
\label{sec:intro}

Neutrinos are among the most abundant particles in the observed Universe. In nature, they are produced by nuclear reactions inside stars, radioactive decays, and cosmic ray interactions. However, due to their tiny interaction cross sections with ordinary matter, their detection is notoriously challenging. In laboratory facilities, neutrino beams are produced via weak decays of hadrons such as pions and kaons, which primarily yield muon neutrinos ($\nu_\mu$) with energies up to~100 GeV~\cite{T2K:2011qtm, DUNE:2020ypp, DUNE:2016hlj, DUNE:2021tad, MicroBooNE:2015bmn, AlvarezGarrote:2024szs, ICARUS:2004wqc}. The most energetic human-made neutrinos have recently been observed at the LHC by the FASER experiment~\cite{Feng:2017uoz, FASER:2019dxq, FASER:2021mtu, FASER:2022hcn} and confirmed afterwards by SND@LHC~\cite{SNDLHC:2023pun}. FASER has detected electron neutrinos ($\nu_e$) and muon neutrinos between \SI{100}{GeV} and several TeV from the decay of forward pions, kaons and charm hadrons produced in LHC collisions~\cite{FASER:2023zcr, FASER:2024hoe}. Tau-neutrinos ($\nu_\tau$) in this energy range, however, are still elusive due to their much lower production rates from the decays of heavy meson ($D_s^\pm$, $B^\pm$, etc.). On the other hand, $\nu_\tau$'s are precisely most-wanted in connection to possible new physics beyond the Standard Model (BSM). 

Recently, there has been renewed interest in a high-energy muon collider (MuCol) \cite{InternationalMuonCollider:2025sys}. The primary motivation for such a machine is to open new territory at the energy frontier (\SI{10}{TeV} partonic center-of-mass energy) in searches for physics beyond the Standard Model. Owing to the muon's short lifetime, muon collider beams decay continuously, leading to an enormous flux of multi-TeV neutrinos with well-understood energy spectra and flavor compositions: $\nu_\mu$ and $\bar\nu_e$ from $\mu^- \to e^- \nu_\mu \bar\nu_e$ decay, $\bar\nu_\mu$ and $\nu_e$ from $\mu^+ \to e^+ \bar\nu_\mu \nu_e$ decay. There are a number of proposals for using these neutrinos for precision measurements of SM parameters and the potential discovery of new physics in this unprecedented energy regime~\cite{King:1997dx, InternationalMuonCollider:2024jyv, Bojorquez-Lopez:2024bsr, deGouvea:2025zfq, Kling:2025zsb, Adhikary:2024tvl}.

In this paper, we present a detailed calculation of muon collider neutrino fluxes and spectra of all flavors using the latest MuCol reference design~\cite{MuCoL:2024oxj, MuCoL:2025quu}. Around the interaction points (IPs), we find a significantly enhanced neutrino flux from the decay of beam muons compared to previous estimates. We pay special attention to subdominant production modes which have not been considered previously: (i) the primary $\mm$ collisions will produce a large number of short-lived particles, which subsequently decay to neutrinos, for instance $\ts, \cc$, etc. In this context, vector boson fusion (VBF) channels such as $\gg \to \tau^+ \tau^-$ have particular importance. (ii) Michel electrons from muon decay interact in the material surrounding the beam pipe, especially in the shielding intended to suppress beam-induced backgrounds. The resulting electromagnetic showers lead to the production of a sizeable number of secondary neutrinos. (iii) Interactions of beam neutrinos produce mesons whose decay leads to a secondary flux of ``neutrino-induced neutrinos.''

Our work is organized as follows. In \cref{sec:production} we discuss the four relevant neutrino sources: $\mu^+\mu^-$ collisions, decays of beam muons, electromagnetic showers induced by Michel electrons, and neutrino-induced neutrinos, one by one.  Adding up these contributions, we predict in \cref{sec:flux} the neutrino flux and event rate in a hypothetical forward neutrino detector. In \cref{sec:physics}, we finally highlight several physics opportunities afforded by muon collider neutrinos. We summarize and conclude in \cref{sec:sum}.

\section{Neutrino Production at a Muon Collider}
\label{sec:production}

\subsection{Neutrinos from $\mu^+\mu^-$ Collision}
\label{sec:collisions}

Although the primary purpose of a muon collider would be the exploration of new physics at the energy frontier, $\mu^+\mu^-$ collisions will also copiously produce SM particles, including neutrinos. In this work we will assume a $\mu^+\mu^-$ collider with a fixed collision energy and a typical integrated luminosity:
\begin{equation}
   \sqrt{s} = \SI{10}{TeV}, \qquad {\cal L} = \SI{10}{ab^{-1}} .
\end{equation}
\begin{figure}[t]
    \centering
    \begin{minipage}[b]{0.29\linewidth}
        \centering
        \includegraphics[width=\linewidth]{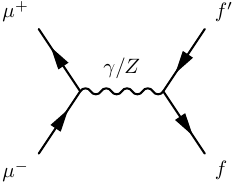}\\[2pt]
        (a)
    \end{minipage}
    \hfill
    \begin{minipage}[b]{0.29\linewidth}
        \centering
        \includegraphics[width=\linewidth]{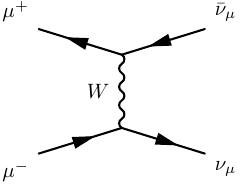}\\[2pt]
        (b)
    \end{minipage}
    \hfill
    \begin{minipage}[b]{0.395\linewidth}
        \centering
        \includegraphics[width=\linewidth]{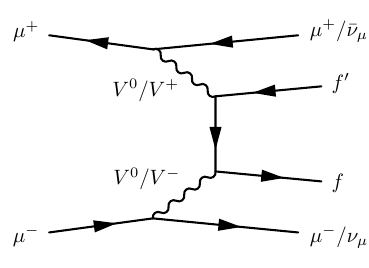}\\[2pt]
        (c)
    \end{minipage}
    \caption{Representative diagrams for the leading order (a, b) and higher-order (c) neutrino production channels in $\mu^+\mu^-$ collisions. $V^0$/$V^\pm$ represent neutral/charged vector bosons and $f$/$f'$ represent final state fermions/anti-fermions.}
    \label{fig:HO_Diagram}
\end{figure}
The cross section for a generic production process $\mu^+ \mu^- \to X$ can be calculated in leading-order perturbation theory by means of standard tools. It is often convenient to write the cross section for $X$ production in terms of a partonic luminosity $\mathcal{L}_{i,j}$ by convolving the partonic cross section $\hat{\sigma}(ij\to X)$ with the appropriate parton distribution functions (PDFs) $f_{i,j}$:
\begin{align}
    \label{eq:gen_mupmum}
    \sigma(\mu^+\mu^-\to X) &= \sum_{i,j} 
        \int_{\tau_0}^1 \! d\tau \, \frac{d\mathcal{L}_{ij}}{d\tau} \hat{\sigma}(ij \to X)
                                                \nonumber
\intertext{where}
    \frac{d\mathcal{L}_{i,j}}{d\tau} &=
        \frac{1}{1 + \delta_{ij}} \int_{\tau}^1 \frac{d\xi}{\xi} \left[ 
            f_{i}\big(\xi,Q^2\big) f_j \Big(\frac{\tau}{\xi},Q^2\Big)
            + (i\leftrightarrow j)
        \right] .
\end{align}
In these expressions, $\tau=\hat s/s$ is the ratio of the partonic squared center-of-mass energy and the collider squared center-of-mass energy, $\tau_{0} = m_X^2/s$ is the production threshold, $Q^2$ is the factorization scale and $f_{i,j}(\xi,Q^2)$ are the PDFs as a function of the momentum fraction $\xi$ and the appropriate scale $Q$. At leading order, $d\mathcal{L}_{\mm}/{d\tau} = \delta(1 - \tau)$. At high energies, multiple splittings need to be taken into account. The resulting potentially large collinear logarithms are resummed into the PDFs by solving the Dokshitzer--Gribov--Lipatov--Altarelli--Parisi (DGLAP) equations \cite{Han:2020uid, Garosi:2023bvq}. In the following, we will make use of both, using the leading-order full calculation and PDF factorization as appropriate.

At leading order, $\nu_e \bar{\nu}_e$ and $\nu_\tau \bar{\nu}_{\tau}$ production in $\mu^+\mu^-$ collisions proceed via an $s$-channel $Z$-boson exchange, as shown in \cref{fig:HO_Diagram}\,(a). $\tau^+\tau^-$ and $c \bar c$ production, followed by $\tau$ and charm decays to neutrinos, also contribute. However, the corresponding cross sections fall as $\sigma \propto 1 / s$, which would not make a substantial contribution at high collision energies. 
A significant effect in these processes is the initial state radiation (ISR), which leads to the ``radiative return'' of the $Z$ boson \cite{Chakrabarty:2014pja}. In contrast, $\nu_\mu \bar{\nu}_\mu$ production is enhanced because it receives an extra contribution from a $t$-channel $W$ exchange as shown in \cref{fig:HO_Diagram}\,(b).

Looking at additional neutrino production channels in $\mu^+ \mu^-$ collisions,  owing to the copious collinear radiation, a high-energy lepton collider is effectively also a vector-boson collider through VBF processes \cite{Han:2020uid}, as depicted in~\cref{fig:HO_Diagram}(c).
Although formally higher order in perturbation theory, the VBF 
processes receive collinear enhancements of the form $\log(Q^2/m^2_\ell)$ for photon-initiated processes, and $\log(Q^2/m^2_W)$ for massive gauge boson processes, where $m_\ell$ and $m_W$ are the charged lepton and $W$ boson masses, respectively. For VBF involving photons, it is immediately clear from the smallness of the lepton masses that the collinear logarithms dominate and require resummation into PDFs. For VBF via $W/Z$ fusion, the factorization scale $Q^2$ has to be much larger than $m_W^2$ for the partonic picture to yield precise results \cite{Dahlen:2025udl}. If $Q^2 \lesssim m_W^2$, the partonic picture for $W/Z$ is invalid and the fixed order  diagrammatic calculations should be used. 

\begin{figure}
    \centering
    \includegraphics[width=0.75\linewidth]{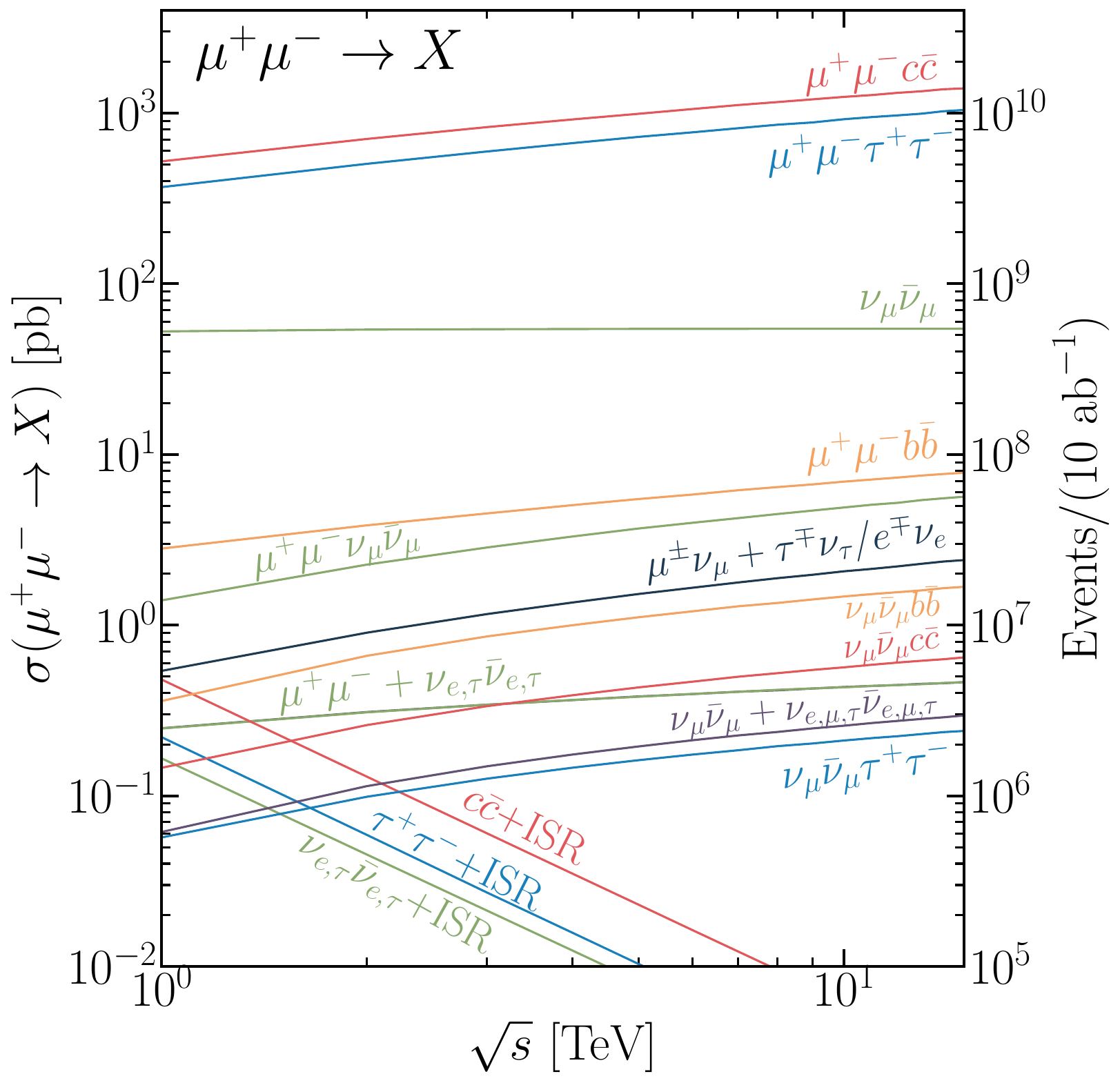}
    \caption{Production cross sections for processes yielding neutrinos in $\mm$ collisions as a function of the collider center-of-mass energy $\sqrt{s}$ with no kinematic cuts on the final state particles. The right axis shows the number of events expected at an integrated luminosity of \SI{10}{ab^{-1}}. } 
    \label{fig:mupmumxsec}
\end{figure}

We consider first the process $\mu^+ \mu^- \to \mu^+\mu^- \tau^+\tau^-$, which turns out to be the dominant source of neutrinos.  At the partonic level, this corresponds to $\gamma\gamma \to \tau^+\tau^-$, and the partonic cross section is given by 
\begin{align}
    \label{eqn:VBS_subprocess}
    \hat\sigma(\gamma\gamma\to \tau^+\tau^-) &=
        \frac{4\pi\alpha^2}{\hat s} \bigg[
            \frac{3 - \beta_\tau^4}{2}
            \ln{\frac{1 + \beta_\tau}{1 - \beta_\tau}}
            - \beta_\tau \big(2 - \beta_\tau^2 \big)
        \bigg] \,, 
\end{align}
with $\beta_\tau \equiv \sqrt{1 - 4 m^2_\tau/\hat s}$, and $\alpha$ being the electromagnetic fine structure constant. Adopting the leading-order splitting $\ell\to \ell\gamma$, which corresponds to the Equivalent Photon Approximation (EPA), the photon PDF as a function of the photon energy fraction $x_\gamma \equiv E_\gamma/E_\ell$ is \cite{Budnev:1975poe} 
\begin{align}
    f_{\gamma/\ell}(x_{\gamma}) &=
        \frac{\alpha}{\pi} \frac{1}{x_{\gamma}} \bigg[
            \bigg( \bar{x}_{\gamma } + \frac{x_{\gamma}^2}{2} \bigg)
            \log\frac{Q_{\max}^2}{Q_{\min}^2}
          - \frac{m_\ell^2 x^2_\gamma}{Q_{\min}^2} \bigg(
                1 - \frac{Q_{\min}^2}{Q_{\max}^2} \bigg)
        \bigg]
    \label{eq:epa}
\intertext{where}
    Q_{\min}^2 &= \frac{m_{\ell}x_{\gamma}^2}{\bar{x}_{\gamma}},
    \qquad
    Q_{\max}^2 = s\bar{x}_{\gamma},
    \qquad
    \bar{x}_\gamma = 1 - x_\gamma \,.
    \label{eq:QminQmax}
\end{align} 
Here we have included the kinematic terms up to the order $x^2_\gamma$.
The collinear enhancement factor $\log(Q_{\max}^2/Q^2_{\min})$ and the infrared divergence $\propto 1/x_\gamma$ are manifest in \cref{eq:epa}. The scales $Q_{\min}$, $Q_{\max}$ given in \cref{eq:QminQmax} apply in the absence of additional kinematic cuts. Using \cref{eqn:VBS_subprocess,eq:epa}, the full cross section $\sigma(\mu^+\mu^- \to \mu^+\mu^- \tau^+\tau^-)$ can now be calculated from \cref{eq:gen_mupmum}.

The EPA offers only a leading-order estimate. For more accurate results, higher-order corrections from multiple collinear splittings need to be included. We have therefore calculated the photon PDF of the muon by solving the DGLAP equations, including all relevant SM particles in the evolution \cite{Han:2020uid, Garosi:2023bvq}. This corresponds to the resummation of the leading collinear logarithms (LL) to all orders in perturbation. The theoretical uncertainty in the partonic treatment can be expressed in terms of an effective factorization scale $Q^2<Q_{\max}^2$ which depends on the kinematics of the hard process.

To validate our treatment, we have performed an independent calculation at fixed order. We use the Monte Carlo package \texttt{Whizard}~\cite{Kilian:2007gr, Moretti:2001zz} to calculate $\mu^+\mu^-\to \mu^+\mu^-\tau^+\tau^-$ including all leading-order diagrams. We pay special attention to the collinear and soft singularities associated with the small lepton masses. Although limited to the leading order, there is no scale uncertainty in this calculation. We find that by choosing a fixed  scale $Q \approx \sqrt{s}/4$ in place of $Q_{\max}$ from \cref{eq:epa}, our cross section from the resummed PDF approach agrees  with the fixed-order calculation within $\mathcal{O}(10\%)$. 

Furthermore, at multi-TeV energies we calculated the resummed PDFs down to very small momentum fractions $x_{\gamma}\gtrsim m_{\tau}^2/s\approx 10^{-8}$. We find that the PDFs do not vary significantly between numerical DGLAP and analytical EPA approaches in this unprecedented kinematic regime, which gives us confidence in our treatment of soft photons. 

The process $\mu^+\mu^-\to \mu^+\mu^-c\bar{c}$, which produces neutrinos through the decays of charm mesons, can be treated in complete analogy to $\mu^+\mu^- \to \mm \tau^+\tau^-$. It differs in the electromagnetic charge and color factors, as well as in mass. However, we note that, even though $\mu^+ \mu^- \to \mu^+\mu^- c\bar{c}$ has a larger cross section, subsequent decay branching fractions and hadronization factors lead to a lower neutrinos yield. Moreover, $\mu^+\mu^-\to \mu^+\mu^- b \bar{b}$ is fully analogous, but the smaller electric charge ($Q_b^4=1/81$) and larger bottom quark mass results in a cross section that is parametrically suppressed.

We have also considered VBF processes involving $W$ bosons \cite{Han:2020uid}, which include $\gamma W^\pm \to \ell^\pm \nu,\ WW\to \nu \bar \nu, c\bar c, b\bar b$. However, owing to the much lower gauge-boson flux compared to photons even at muon collider energies, the cross section contributions for these processes are smaller than the photon-induced processes by two to three orders of magnitude. 

We present the cross sections for different neutrino production channels in \cref{fig:mupmumxsec}, including both direct production in the primary interactions and secondary production through decays of $\tau$, charm and bottom. The cross sections shown here correspond to a leading-order calculation in \texttt{Whizard}, but we have individually cross-checked them against the PDF approach wherever appropriate.
We see that the $\gg \to \ts$ cross section reaches $(0.4-1)$~nb within the energy range $\sqrt s = 1-10$~TeV. Scaled by electric charge and a color factor, and taking into account the reduced running mass of the charm charm quark ($m_c \approx \SI{0.8}{GeV}$ at $\sqrt{\hat{s}} = \SI{10}{GeV}$) the cross section of $\gg \to c\bar c $ is larger by approximately 30\%, reaching values of about $(0.5-1.3)$~nb in the same energy range. The $\gg \to b\bar b$ cross section, which is lower by more than two orders of magnitude, is comparable to the contributions from charged current $WW$ fusion, of order $(2-10)$~pb.
This is notably lower than the cross section of the leading-order process $\mm \to \nu_\mu \bar \nu_\mu$, mediated via $t$-channel $W$ exchange, for which logarithmic enhancement at high energies leads to $\sigma \sim \SI{50}{pb}$. 

The smallest cross sections appearing in \cref{fig:mupmumxsec} belong to the $s$-channel production processes like $\mm\to \nu_e \bar{\nu}_e,\ \nu_\tau \bar{\nu}_\tau$ and $\tau^+\tau^-$ and $c\bar c$.
In computing these cross sections, we have been careful to include ISR effects which lead to an enhancement by a factor 2--7, depending on the channel. 
The cross sections scale approximately as $\sigma \sim \SI{200}{fb} / (s/\si{TeV^2})$, which makes them less important at high collision energies. 

\begin{figure}
    \centering
    \includegraphics[width=1\linewidth]{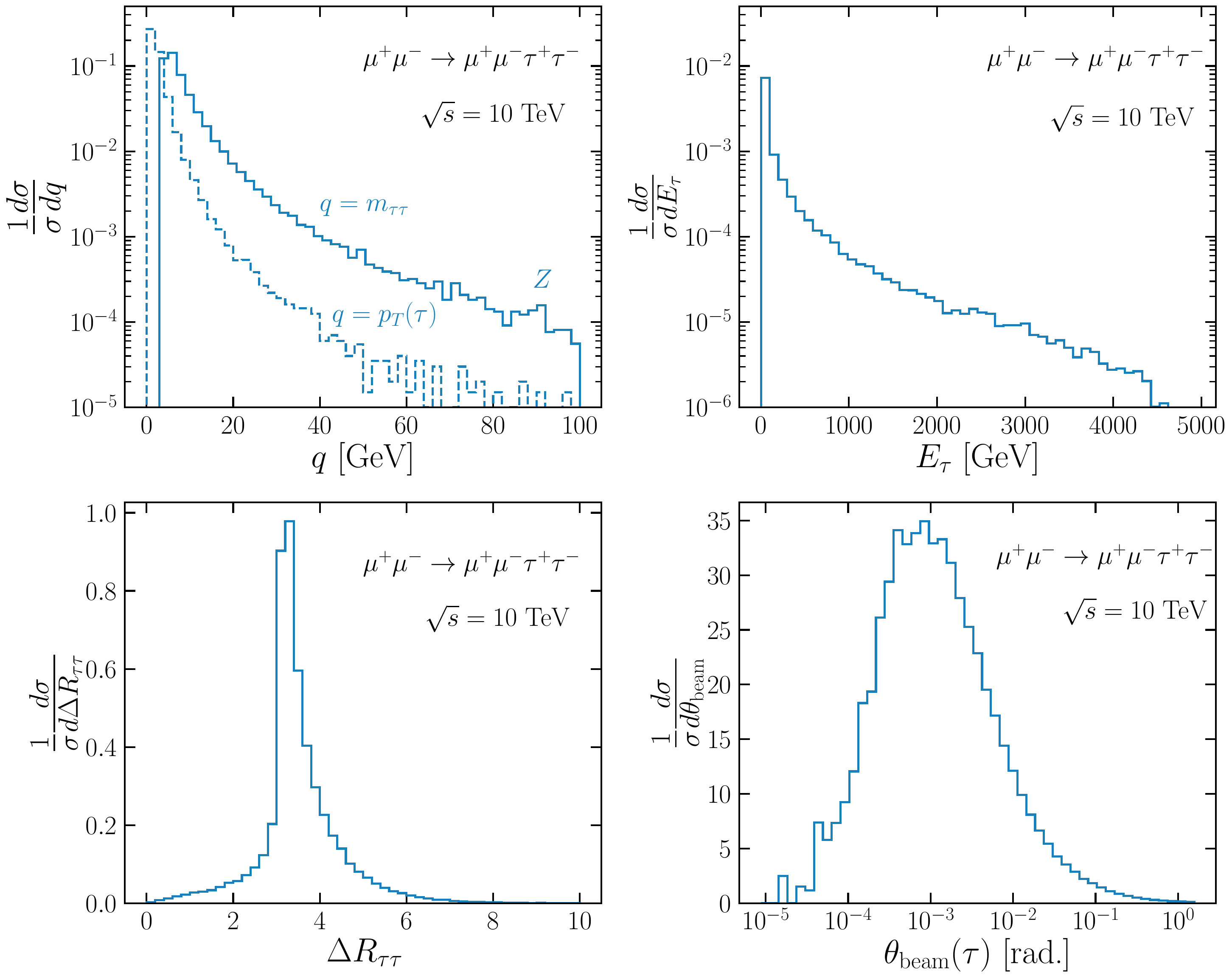}
    \caption{Kinematic Distributions of the  $\tau^+\tau^-$ pair, for the invariant mass and transverse momentum distributions (top left), the tau energy $E_{\tau}$ (top right), the angular separation $\Delta R_{\tau\tau}$ (bottom left) and the lab frame beam angle $\theta_{\text{beam}}$ (bottom right).} 
    \label{fig:kindists}
\end{figure}

Turning from cross sections to event rates, the leading process $\gg\to \ts$ is expected to produce about $10^{10}$ $\nu_\tau$ for an integrated luminosity of \SI{10}{ab^{-1}} (right axis in \cref{fig:mupmumxsec}).
It is interesting to note that because of this process, the neutrino flux from $\mu^+\mu^-$ collisions is dominated by $\nu_\tau$, with other neutrino flavors being subdominant.
The second-largest source of neutrinos (again $\nu_\tau$) is from $\mu^+\mu^- \to \mu^+\mu^- c\bar{c}$, followed by hadronization into $D_s$ and decays via $D_s \to \tau\nu_\tau$. 
We have simulated the hadronization and decay using \texttt{Pythia}~8.3~\cite{Sjostrand:2014zea, Bierlich:2022pfr}, including  polarization effects. We find that the efficiency factor from the charm fragmentation and $D_s$ decay branching fraction is of order 1\%.

Besides the total production cross sections, it is also important to understand the final state kinematics of neutrino production in $\mu^+\mu^-$ collisions.
Several (normalized) kinematic distributions are displayed in \cref{fig:kindists} for the process $\mu^+\mu^- \to \mu^+\mu^-\tau^+\tau^-$.
In the top left panel we show the $\ts$ invariant mass distribution, $m_{\tau\tau}$ (solid blue histogram). Clearly visible are the kinematic threshold at $2m_\tau$, and the drop-off $\propto 1/m^2_{\tau\tau}$ towards higher invariant masses, characteristic of VBF processes. 
There is a notable but insignificant contribution from $Z\to \ts$ near $m_Z$.
Also shown in the top left panel of \cref{fig:kindists} is the $\tau$ transverse momentum distribution, $p_T(\tau)$ (dashed blue), which is roughly correlated as $\sigma(p_T)\sim \sigma(m_{\tau\tau}/2)$.  
The tau lepton energy distribution is shown in the top right panel of \cref{fig:kindists}. As can be seen, although the energy of $\tau$ is sharply peaked near threshold, the histogram extends to the multi-TeV range. This implies the production of very high energy neutrinos in the decays of these $\tau$s.

In the bottom left panel of \cref{fig:kindists}, we show the angular separation between the $\tau^+$ and $\tau^-$, $\Delta R \equiv \sqrt{\Delta\phi^2+\Delta\eta^2}$, where $\phi$ is the azimuthal angle and $\eta$ the pseudorapidity. We see that the distribution sharply peaks at $\Delta R = \pi$, corresponding to back-to-back $\tau$s. This corresponds to the expected kinematics for the partonic process $\gg \to \ts$. 
Finally, in the bottom right panel of \cref{fig:kindists}, we plot the emission angle of the $\tau$ in the lab frame with respect to the beam axis. We see that most of the events occur in a rather small angular interval, $\theta_{\rm beam} \sim m_\tau/E_\mu \simeq 1$~mrad. This observation is relevant in chosing the optimal placement of a neutrino detector.

As a final remark, it is well known that a high-energy lepton collider will also produce a large number of soft hadrons through photon-induced $\gg \to \text{hadrons}$ processes. It was estimated in Ref.~\cite{Han:2021kes} that the cross section for photon-induced hadron production at a \SI{10}{TeV} muon collider may reach about \SI{100}{nb}, two orders of magnitude above that of $\gg \to \ts$. However, these soft hadrons would create only low-energy neutrinos, leading at best to background noise in a detector optimized for high energies. Furthermore, even though a large number of charged pions and kaons are produced, they are rather long-lived in the collider environment, and only a small fraction will decay to neutrinos before interacting. The corresponding probability to decay is approximately $P \approx L/c\tau\gamma$. 
The relevant length scale is the distance to the hadronic calorimeter in which these particles will interact, or the distance to the first downstream magnets which deflect them towards absorbers. Therefore, $L \sim \text{meters}$. Conservatively considering an energy $E \sim \SI{100}{GeV}$, we obtain decay probabilities of approximately $10^{-4}$ for the charged pions and kaons, which renders their contribution to the neutrino flux negligible compared to the contribution from prompt particle decays.

\subsection{Neutrinos from Beam Muon Decays}
\label{sec:beam}

Any circular muon collider must include a muon storage ring which delivers the beam to the collision points. Because of their short lifetime, the stored muons decay continuously and produce a large number of neutrinos. In this study, we consider the neutrino fluxes expected from a muon collider operating at a center-of-mass energy of \SI{10}{TeV}. The proposed design features a \SI{10}{km} storage ring into which approximately $10^{13}$ muons are injected every second~\cite{InternationalMuonCollider:2025sys}. This leads to an expected neutrino production rate of about $10^9$ decays per meter per second. Although muon decays occur throughout the ring, the most intense neutrino beam is generated in the straight sections near the interaction points (IPs). The exact beam intensity depends strongly on the machine geometry, particularly the length of the straight sections and the beam optics.

Preliminary estimates of the neutrino flux from beam muon decay were presented in Refs.~\cite{InternationalMuonCollider:2024jyv, InternationalMuonCollider:2025sys} and subsequently utilized in the phenomenological studies of Ref.~\cite{Adhikary:2024tvl, Kling:2025zsb}. In the absence of a detailed design of the interaction region, these studies conservatively assumed the length of the straight section to be $L = \SI{10}{m}$, roughly corresponding to the size of the main detector. This leads to a flux of $\phi_\nu \approx 10^{10}$ neutrinos per second. Furthermore, an angular divergence of the muon beam of \SI{0.6}{mrad} was assumed, which fixed the angular size of this neutrino beam.

\begin{figure}
    \centering
    \includegraphics[width=0.98\linewidth]{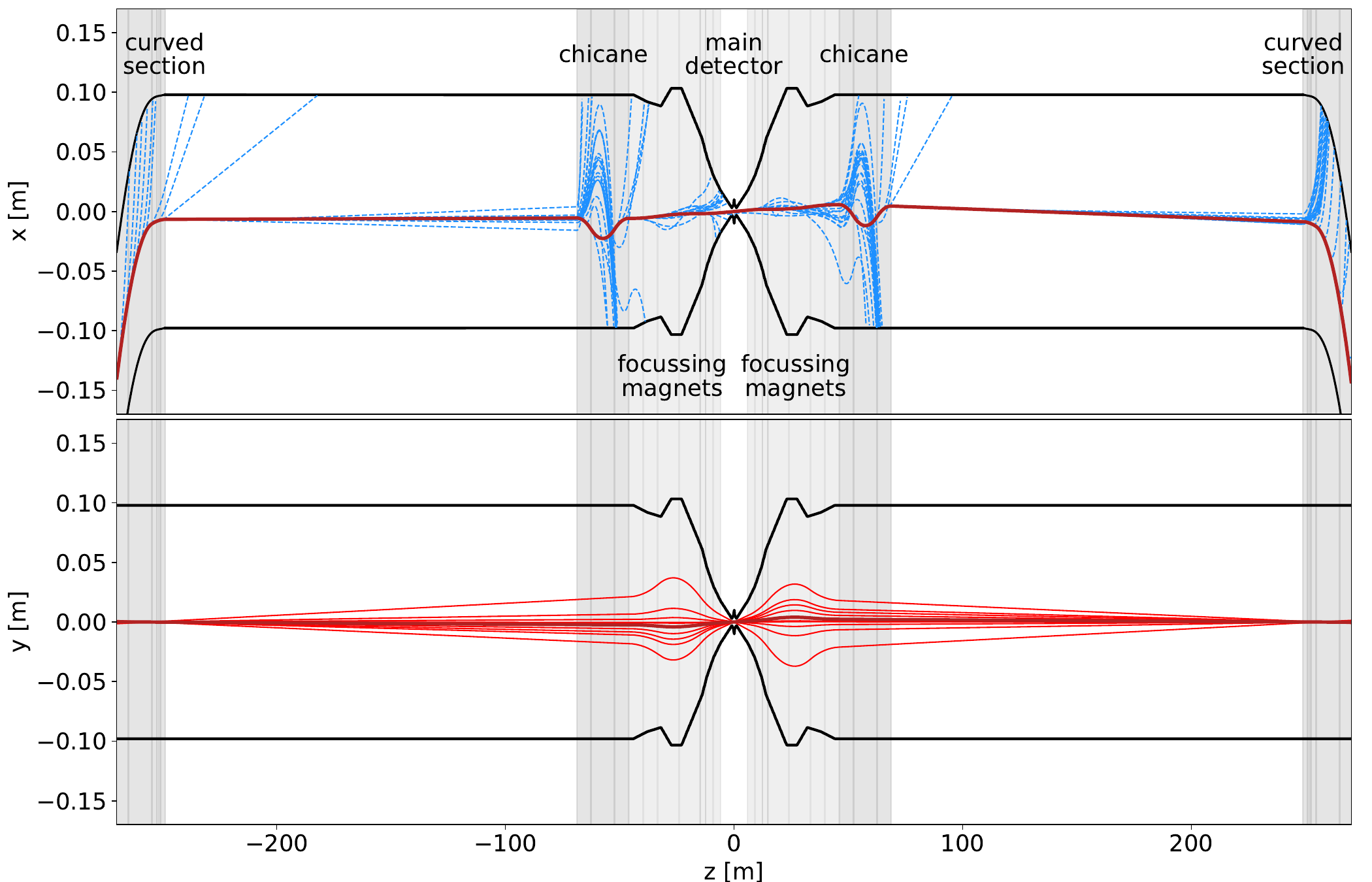}
    \caption{The geometry of interaction region is shown in the horizontal (top) and vertical (bottom) planes. The shaded areas correspond to the quadrupole (light gray) and dipole (dark gray) magnets. The solid dark red line shows one example beam muon trajectory. The dashed blue lines in the top panel show the trajectories of electrons from their production until they hit the absorber surrounding the beam pipe.
    The red lines in the bottom panel show additional example trajectories of muons in the beam sampled according to a divergence of \SI{0.6}{mrad} at the IP.
    Black lines indicate the inner aperture of the beam pipe.
    }
    \label{fig:trajectory}
\end{figure}

In this work, we refine the above estimate. We consider the design of the interaction region presented in Refs.~\cite{MuCoL:2025quu, InternationalMuonCollider:2025sys} (corresponding to the v0.8 configuration). This layout is illustrated in \cref{fig:trajectory}, where we show cross sectional views in the horizontal and vertical planes. Located at $x=y=z=0$ is the primary IP. It is surrounded by the main detector, which extends to $|z|=6$. Located between $z=6$ and \SI{46}{m} downstream of the IP are quadrupole magnets which act as beam lenses. Together with additional quadrupole magnets further downstream, these allow a beam angular divergence of \SI{0.6}{mrad} at the IP and ensure collimated parallel beam trajectories at the entrance of the curved part of the collider ring. Placed after these quadrupole magnets is a series of dipole magnets, the chicane, which bends the muon beam and then returns it to its original trajectory while electrons from muon decay are filtered out, thereby reducing the radiation load on the experiments. The curved part of the ring starts at $|z| \approx \SI{250}{m}$. It contains a lattice of dipole magnets to bend the beam, and additional focusing quadrupole magnets. Here we consider the design presented in Ref.~\cite{Skoufaris:2023jnu} (with the configuration files of Ref.~\cite{acc-models}).

To protect the main detector from intense beam induced backgrounds, mainly electrons from muon decay, the detector is shielded by the \textit{nozzle}: a thick layer of tungsten absorber. Similarly, magnets and other downstream infrastructure elements are shielded by a few cm thick absorption layer. Of primary importance for this study is the inner aperture of the beam pipe, which is shown as a black line in \cref{fig:trajectory}. Here we use the design of the nozzle presented in Ref.~\cite{MAIA:2025hzm}. For the downstream part, we follow the information provided in Ref.~\cite{MuCoL:2024oxj, MuCoL:2025quu} and assume, for simplicity, that the radius of the beam pipe in the downstream part of the straight section remains similar to that of the chicane. 

Having defined the geometry, we can simulate a set of trajectories. As an initial condition, we require that all muons have a momentum of $E_\text{beam} = \SI{5}{TeV}$ and that their trajectories converge at the IP. We further sample the beam's transverse momentum from a gaussian distribution with a width of $\SI{0.6}{mrad} \times E_\text{beam}$. We then propagate the muons through the beam pipe and magnets using the implementation obtained in Ref.~\cite{Kling:2021gos}. We utilize the \texttt{BDSIM Quadrupole} and \texttt{BDSIM Dipole Rodrigues} first order matrix tracking algorithms that are described in the \texttt{BDSIM} documentation~\cite{Nevay:2018zhp, BDSIM}. The simulation was validated in Ref.~\cite{Kling:2021gos} against the full \texttt{BDSIM} prediction and great agreement of $\mathcal{O}(\SI{10}{\mu m})$ and better was found. Several example trajectories are shown as red lines in the bottom panel of  \cref{fig:trajectory}.

To obtain the neutrino beam, we simulate 1000 distinct trajectories and sample the decays of muons into neutrinos roughly every meter along these trajectory within the range $|z| < \SI{270}{m}$. The decay itself is simulated in \texttt{Pythia}~8.3 in the muon rest frame, and the resulting neutrinos are then boosted into the laboratory frame. The resulting neutrino flux is found to be about $10^{12}$ neutrinos per second or $10^{19}$ neutrinos per year in the forward direction. This is roughly a factor of 120 larger than the previous estimate~\cite{InternationalMuonCollider:2025sys}. There are two main reasons for this: firstly, we consider neutrino decays not only in the $L \approx \SI{10}{m}$ central part of the straight section, but from its whole length $L = \SI{500}{m}$. Secondly, the beam divergence decreases away from the IP, leading to a more collimated and therefore more intense neutrino beam.

We note that the expected neutrino flux depends on the design of the interaction region. In this work, we use design version~0.8 (v0.8), as presented in the input to the 2025 European Strategy Update. However, the interaction region is still under active development, and the final design has not yet been established. For example, compared to the earlier v0.6~\cite{Skoufaris:2022wwg}, v0.8 approximately doubles the length of the straight section and places the chicane downstream of the quadrupole magnets. To assess the impact of these design changes, we also simulated v0.6 and found that the neutrino flux differs by approximately a factor of two, consistent with the change in the straight-section length. This comparison illustrates that the predicted flux is sensitive to the machine layout and that the final flux is therefore expected to change by an $\mathcal{O}(1)$ factor as the interaction-region design continues to evolve. For the interested reader, we provide a more detailed presentation of the comparison between design versions in Appendix \ref{App:Flux_Comparisions}. 

While muon decays only produce muon neutrinos and electron anti-neutrinos, other neutrino flavors can arise due to neutrino oscillations. Later we will consider a detector placed $L = \SI{500}{m}$ downstream of the IP. The oscillation probability for neutrinos with energies of $E \sim \SI{1}{TeV}$ is given by $P_{\alpha \to \beta} \sim \sin^2( \Delta m_\text{atm}^2 L / (4 E)) \sim 10^{-12}$, where $\Delta m_\text{atm}^2 = \SI{2.5e-3}{eV^2}$ is the atmospheric mass splitting. Due to the large flux of neutrinos from muon decays, even such a small oscillation probability will lead to a sizable flux of $\nu_e$, $\bar\nu_\mu$, $\nu_\tau$ and $\bar\nu_\tau$ of about $10^7$ neutrinos per year at a distance of \SI{500}{m} distance from the primary collision point. When providing numerical results, we will consider the full three-flavor oscillation probability.

\subsection{Neutrinos from Electron-Induced Showers}
\label{sec:shower}
In addition to neutrinos, decays of beam muons also produce high-energy electrons. These electrons then propagate through the muon collider's beam pipe and magnets before they start electromagnetic showers in the absorbers surrounding the beam pipe. Following the same arguments as in \cref{sec:beam}, we expect the production of $10^9$ such electrons per second and per meter. Along the \SI{500}{m} long straight section, this corresponds to about $10^{19}$ showers per year. Due to this large number of showers, even relatively rare tau-lepton or $D$-meson production processes inside the shower lead to a substantial flux of neutrinos, as we will see in the following.

As in the previous section, we sample the decays of muons into electrons roughly every meter along the muon trajectory within the range $|z| < \SI{250}{m}$. We obtain the energy distribution of the decay electrons in the muon's rest frame from the Michel parameterization, $d\Gamma/dx_e \sim x_e^2(3-2x_e)$, with $x_e \equiv E_e/m_\mu$ being the fraction of muon mass carried away by the Michel electron. We assume an isotropic angular distribution. After boosting the electrons into the laboratory frame, they are propagated through the magnetic fields. Unlike muons, the electrons experience a sizable energy loss due to synchrotron radiation. The rate of energy loss per unit distance traveled by a relativistic electron can be estimated as~\cite{Jackson:1998nia}
\begin{align}
    \frac{dE}{ds} = \frac{e^4 B^2 E_e^2}{6 \pi m_e^4}
        = \SI{1.2}{GeV/m} \times
          \bigg( \frac{E_e}{\si{TeV}} \bigg)^2
          \bigg( \frac{B}{\si{T}} \bigg)^2 \,, 
\end{align}
where $e$ is the unit electric charge, $B$ is the magnetic field strength and $E_e$ is the total electron energy. This energy loss is particularly significant in the chicanes, which are about \SI{20}{m} long, with a field strength of about \SI{10}{T}. At this field strength, a \SI{3}{TeV} electron will lose almost \SI{1}{TeV} within the first meter. Taking into account both deflection and energy loss through synchrotron radiation, we propagate the electrons through the magnets until they hit the beam pipe aperture, where they initiate an electromagnetic shower. Since the absorbers surrounding the pipe will be made of dense tungsten alloys with correspondingly small radiation lengths, we assume the entire shower to develop inside the shielding~\cite{MuCoL:2024oxj}.

In the upper panel of \cref{fig:trajectory}, we show a representative muon trajectory (solid red) together with possible trajectories of electrons produced in the decay of this muon along its trajectory (dashed blue). Electrons produced between the dipole magnets move almost parallel to the muon beam and can therefore propagate over a sizable distance. However, once they enter the dipole magnets, either in the chicane or in the curved sections, they rapidly lose energy and are strongly deflected. In contrast, electrons produced inside the focusing quadrupole magnets upstream of the experiments experience a weaker energy loss and deflection before they hit the shielding close to the main detector. Their typical angle at the onset of the shower is about one order of magnitude larger than the muon beam divergence. This will lead to a bigger angular spread of neutrinos produced in showers compared to neutrinos produced in beam muon decays. We note that photons from synchrotron radiation typically have energies $E_\gamma \sim \frac{3}{2} \frac{e B E_e^2}{m_e^3} \sim \text{few GeV}$ for $E_e \sim \si{TeV}$ and $B \sim \SI{10}{T}$, with higher energy photons being exponentially suppressed, and therefore do not lead to the production of high-energy neutrinos \cite{Jackson:1998nia}.  

To simulate the electromagnetic showers initiated by the electron, we employ a shower simulation originally built for Ref.~\cite{Kling:2025lnt} as inspired by Ref.~\cite{Salvat:2019pen}. The algorithm models both $\gamma \to e^+e^-$ pair production and Bremsstrahlung $e^\pm \to e^\pm + \gamma$ following Ref.~\cite{ParticleDataGroup:2024cfk, Klein:1998du, PhysRevD.25.1291}. 
For the latter, we account for the Landau-Migdal-Pomeranchuk effect~\cite{Landau:1953ivy, Landau:1953gr, PhysRev.103.1811} to avoid the singularity of the Bremsstrahlung cross section at $x_\gamma=0$, where $x_\gamma$ is the fraction of the electron energy transferred to the photon. Transverse momentum smearing is added following a distribution function of $1/(p_T^2 + b^2)$ for $0 < p_T < x_\gamma E_e \theta_{\rm max}$. Here $b = x_\gamma^2 m_e^2$ for Bremsstrahlung and $b = m_e^2$ for pair production, and $\theta_{max} = 0.1$ to ensure that the small angle approximation is valid. The algorithm is recursively applied on the additional photons and $e^\pm$ until the energy of the produced particles reaches $E= 10$ GeV, which we take as a minimum energy below which the contributions to the flux become negligible. We have checked that the algorithm reproduces the known shower distributions \cite{ParticleDataGroup:2024cfk}.

For our purposes, we have extended the shower simulation from Ref.~\cite{Kling:2025lnt} by introducing rare interactions that result in neutrino production. In particular, we have added the following processes: 
\begin{description}
 \item [Tau production in photon conversion ($\gamma N \to \tau^+\tau^- N$).] Photons in the shower can convert into fermion pairs through the pair-production process $\gamma N \to \bar f f N$. In the vast majority of cases, the photon converts into an electron-positron pair; however, more rarely, it can also produce heavier fermions. For neutrino production, the most relevant channel is tau-pair production. The photon-to-fermion conversion cross section is described by the Bethe--Heitler formula \cite{Bethe:1934za,Bethe:1954zz, Davies:1954zz}
 \begin{align}
     \sigma(\gamma N \to \bar f f N) \simeq
         \frac{28}{9} \frac{Z^2 \alpha^3 N_c Q_f^4}{m_f^2}
         \log\bigg(\frac{2 E_\gamma}{m_f}\bigg)
         \bigg( 1 - \frac{4m_f^2}{E_\gamma^2} \bigg)^{1/2}\,,
 \end{align}
 where $Z$ is the atomic number of the nucleus, $N_c$ is the color factor, $E_\gamma$ is the photon energy, $Q_f$ is the electric charge of the fermion $f$, and $m_f$ is its mass. The last factor accounts for the kinematic suppression near threshold. From this expression, we can estimate the probability of producing a $\tau^+\tau^-$ pair as
 \begin{equation}
     P \approx \frac{m_e^2}{m_\tau^2}
         \frac{\log(2 E_\gamma /m_\tau )}
              {\log(2 E_\gamma /m_e )}
        \bigg( 1 - \frac{4m_\tau^2}{E_\gamma^2} \bigg)^{1/2}\,.
     \label{BH-Formula}
 \end{equation}
 Given an estimated $10^{19}$ showers per year along the straight section, we therefore expect $\gtrsim 10^{12}$ $\tau^+\tau^-$ pairs annually, and the same number of $\tau$ neutrinos. We simulate the $\tau$ kinematics in the same manner as for electrons.
 The $\tau$s are then decayed in their rest frame, and the resulting neutrinos are subsequently boosted to the laboratory frame. We include the hadronic decay channels $\tau^- \to \nu_\tau \pi^-$ (11\%), $\tau^- \to \nu_\tau \rho^- (\to \pi^- \pi^0)$ (25\%), and $\tau^- \to \nu_\tau a_1^- (\to 3\pi)$ (18\%) \cite{ParticleDataGroup:2024cfk}, which yield monochromatic $\nu_\tau$ with energy $E_\nu = (m_\tau^2 - m_h^2)/2m_\tau$ (where $m_h$ is the hadron mass), as well as the leptonic channels $\tau^- \to \nu_\tau \bar \nu_\ell \ell^-$. Defining $x_\nu = 2E_\nu/m_\tau$ and $x_{\bar\nu} = 2E_{\bar\nu}/m_\tau$, the energy distribution of the $\nu_\tau$ is given by $d\Gamma/dx_\nu \sim 2x_\nu^2(3-2x_\nu)$, while that of the $\bar\nu_\ell$ follows $d\Gamma/dx_{\bar\nu} \sim 12 x_{\bar\nu}^2 (1-x_{\bar\nu})$. The corresponding $\tau^+$ decay channels are treated as analogously. 
 
 \item [D-meson production in photon conversion ($\gamma N \to \bar{c} c N$).] 
 Photons may also convert to $c\bar{c}$ pairs in matter. The cross section for this process is again given by \cref{BH-Formula}, with the appropriate color and charge factors. We hadronize the charm quark assuming that contributions close to the threshold dominate, and that therefore the charm quarks always hadronize into $D$ meson pairs in the collision's center of mass frame. We include the hadronization channels that are most important for neutrino production, namely $c\to D^0 ~(61\%)$, $c\to D^+ ~(24\%)$, $c\to D_s^+ ~(8.1\%)$, and $c\to \Lambda_c^+ ~(6.1\%)$, with the same fractions for the $\bar{c}\to \bar{D}^0$, $D^-$, $D_s^-$, and $\Lambda_c^-$, respectively \cite{Lisovyi:2015uqa, Bhattacharya:2023zei}. For the neutrinos produced in the subsequent meson decays, we determine the energy distribution in the meson rest frame using \texttt{Pythia}~8.3, and we then boost the resulting neutrino momenta to the laboratory frame. We note that, in comparison to the di-tau production mode, these channels increase our electron and muon-flavor neutrinos, but the small fragmentation fraction for $c \to D_{s}^{\pm}$, together with the small $\operatorname{BR}(D_{s} \to \tau \nu_\tau) \sim 5\%$ \cite{ParticleDataGroup:2024cfk}, leads to only a small fraction of $\nu_{\tau}$ production from this channel.

 \item [D-meson production in photo-nuclear interactions ($\gamma  g \to c \bar{c}  X$).]  Charm quarks may also be produced incoherently via the photon--gluon fusion (PGF) process $\gamma + g \to c \bar{c} + X$. The cross section for this process differs from the one for photon conversion to charm quarks by a factor $\simeq \alpha_s^2 A / (Z^2 \alpha^2)$, with $\alpha_s$ the strong coupling constant, $A$ the mass number of the target nucleus, and $Z$ its charge. For tungsten, this factor is $\sim 7$, implying an appreciable enhancement. We estimate the total cross section for PGF $\sigma_{\gamma g}$ as a function of the photon energy $E_\gamma$ by colliding fixed energy photons with protons at rest and simulating the subprocess $\gamma g\to c\bar{c}$ in \texttt{Pythia}~8.3. Assuming that coherent photon conversion to an electron-positron pair is the dominant process, we estimate the probability $P$ of charm quark production via PGF as 
 \begin{equation}P \approx \frac{A\sigma_{\gamma g}}{\sigma(\gamma N\to e^+ e^- N)}.
 \end{equation}
 After hadronizing the $c$ quarks to $D$ mesons using the same fragmentation fractions as described above, we model the distribution of $D$ meson momentum fraction $z\equiv E_D/E_{\gamma}$ and $p_T$ via the ansatz \cite{PhysRevLett.62.513, Filipponi:1997vf}
 \begin{align}
    \frac{d^2\sigma}{dz\,dp_{T}^2} \sim (1 + a \, z)(1 - z)^n
    e^{-b \, p_T^2}\,,
 \end{align}
 where we take $a \approx 14$, $n \approx 3.9$ and $b\approx 1$. The $D$ mesons are then decayed into neutrinos as described above.
\end{description}

To obtain the neutrino flux, we sample all above-mentioned interactions at all stages of the shower. Each tau lepton and $D$ meson is decayed ten times to increase statistics, and the resulting neutrinos are recorded. 

We investigated several additional neutrino production processes in electromagnetic showers, but found them to be negligible. In particular, we considered the decays of muons and light hadrons, whose production probability in the shower scales roughly with the inverse mass squared, $P \propto m^{-2}$. Muons, pions, and kaons are thus produced more abundantly than the taus and $D$-mesons discussed above. However, they are unlikely to decay before either reaching the detector or undergoing a hadronic interaction. The decay probability of a relativistic particle over a distance $L$ is approximately $P \approx L / c\tau\gamma$, where $\tau$ is the particle's lifetime and $\gamma$ its boost factor. For muons, the relevant scale is the distance between the shower and the detector, $L \sim \SI{100}{m}$. For hadrons, however, the relevant scale is the nuclear interaction length, $L \sim \lambda_\text{int}$. Taking $\lambda_\text{int} \sim \SI{10}{cm}$ in tungsten and a representative energy $E \sim \SI{100}{GeV}$, we obtain decay probabilities of approximately $10^{-4}$, $10^{-5}$, $10^{-5}$, $1$, and $1$ for $\mu$, $\pi$, $K$, $\tau$, and $D$ mesons, respectively. Combining these decay probabilities with the corresponding production probability gives relative neutrino contributions of $0.05$, $0.003$, $0.0004$, $1$, and $1$. We therefore conclude that promptly decaying particles, namely taus and charmed mesons, dominate the neutrino yield.

In addition, we considered the coherent lepton trident production process $e N \to e \tau^+ \tau^- N$ as well as charm production via the deep-inelastic scattering process $e N \to e c \bar{c} + X$. However, the associated rate is suppressed by one additional power of $\alpha$ relative to the pair production process $\gamma N \to \tau^+ \tau^- N$ and photo-nuclear interactions $\gamma N \to c \bar{c} + X$. We also investigated $Z$-boson mediated positron annihilation with atomic electrons into neutrinos, $e^+ e^- \to \bar\nu \nu$. This process has a cross section $\sigma = G_F^2 s / 48\pi$. The relative probability for this process scales as $P \sim G_F^2 E_e m_e^3 \alpha^{-3} Z^{-2} \sim 10^{-17}$, rendering its contribution completely negligible.

To end this section, let us re-emphasize that the neutrino flux estimated here depend on the design of the interaction region. To quantify this sensitivity, we have repeated the calculation for the earlier design version~0.6 (v0.6), which featured a roughly two times shorter straight section and did not include the chicanes. We find that the predicted neutrino flux in v0.6 is approximately a factor of five larger than in v0.8, largely due to the absence of the chicane. This comparison again suggests that the predicted event rates may change by an $\mathcal{O}(1)$ factor as the interaction-region layout continues to evolve.  

\subsection{Neutrinos from Neutrino-Induced Showers}
\label{sec:secondary-nu}

Similar to neutrino production in electron-induced showers, also neutrino interactions can produce secondary neutrinos, in particular through charm decays. The starting point for this production mode is the large flux of $\sim 10^{19}$ neutrinos per year that are produced in muon decays and can interact in the rock between the storage ring and the forward detector. The probability for a neutrino to interact while traveling a distance $L$ through rock of mass density $\rho$ is \cite{FASER:2019dxq}
 \begin{align}
     P_\text{int} \approx \num{2.1e-7}
         \times \bigg( \frac{E}{\si{TeV}} \bigg)
                \bigg( \frac{L}{\SI{200}{m}}\bigg)
                \bigg( \frac{\rho}{\SI{2.6}{g/cm^2}} \bigg)\,,
    \label{eq:nu-int-prob}
 \end{align}
 where we take $\rho \SI{2.6}{g/cm^2}$ as the density of standard rock~\cite{ParticleDataGroup:2024cfk}. 
 
 Among the interaction products, light hadrons and muons are long-lived and hence more likely to interact than to decay into neutrinos. Secondary high-energy neutrinos are therefore primarily produced in charm decays. The probability for a TeV-energy neutrino to produce charm is about 10\%. Together with \cref{eq:nu-int-prob}, this leads to a secondary neutrino production probability $\sim 10^{-8}$, which is similar to the charm production probability in photon conversion or photo-nuclear interactions. For the numerical results presented below, we have simulated the kinematics deep-inelastic neutrino scattering and subsequent charm decays with \texttt{Pythia}~8.3, taking the total interaction cross section from \texttt{GENIE~3}~\cite{Andreopoulos:2009rq} as obtained in Ref.~\cite{FASER:2024ykc}.

\section{Neutrino Fluxes and Event Rates in a Forward Neutrino Detector}
\label{sec:flux}

\begin{figure}[!htbp]
    \centering
    \includegraphics[width=1\textwidth]{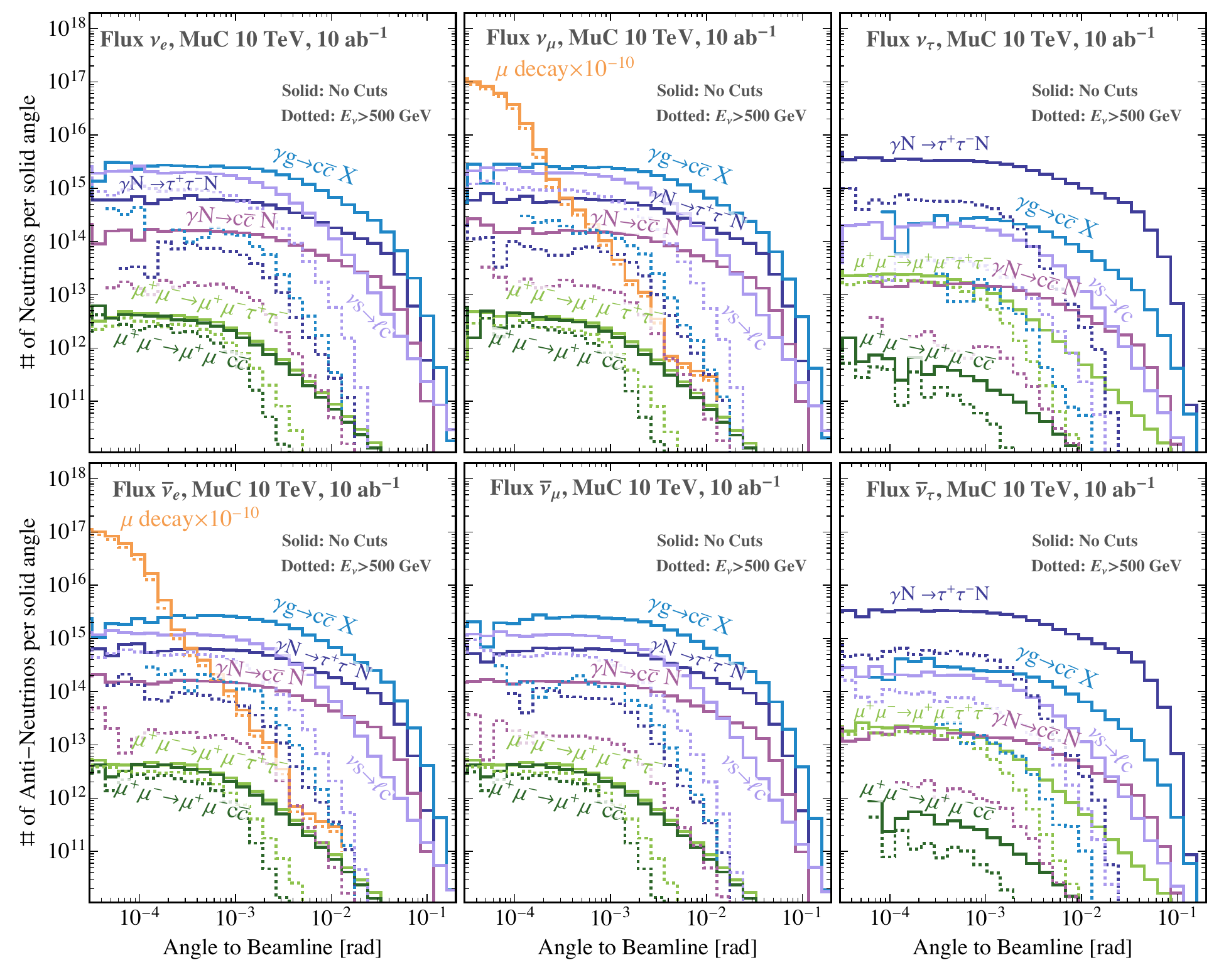}
    \caption{Fluxes of neutrinos (upper panels) and anti-neutrinos (lower panels)  produced in the straight sections of a \SI{10}{TeV} muon collider as a function of the emission angle. We show separately the contributions from beam muon decay, showers and $\mu^+\mu^-$ collisions. We have assumed 10 years of operation, corresponding to $10^{-10}$ of  integrated luminosity.
    }
    \label{fig:FluxesAngle}
\end{figure}

Having identified and simulated all the relevant neutrino production modes, we are now ready to compare the neutrino fluxes. The largest neutrino flux contribution is provided by muon decays, which produce $\mathcal{O}(10^{19})$ muon neutrinos and electron anti-neutrinos per year in the straight section. The second largest contribution is provided by electromagnetic showers initiated by electrons from beam decay which mainly proceeds through tau-lepton and charm production. This contribution is suppressed by roughly a factor $m_e^2/m_{\tau/c}^2 \sim 10^{-7}$, and therefore produces $\mathcal{O}(10^{12})$ neutrinos and anti-neutrinos of all flavors per year in the straight section. A similar contribution is provided by interactions of the neutrino beam in the the rock, producing $\mathcal{O}(10^{11})$ neutrinos and anti-neutrinos of all flavors per year. The smallest contribution is due to the $\mathcal{O}(10^{9})$ tau-lepton and $D$-meson pairs produced in the primary collisions (assuming an integrated luminosity of $\mathcal{L}=1~$ab$^{-1}$) which lead to 
a similar number of neutrinos and anti-neutrinos of all flavors. 

Using the simulations described in the previous sections, we obtain the (anti-)neutrino fluxes of all flavors for a \SI{10}{TeV} muon collider. We have assumed 10 years of running at nominal beam intensity; for neutrinos originating from $\mu^+\mu^-$ collisions have have assumed an integrated luminosity of $10~$ab$^{-1}$, based on the design in Refs.~\cite{MuCoL:2025quu, InternationalMuonCollider:2025sys}. The neutrino fluxes per solid angle are shown in~\cref{fig:FluxesAngle}, where the distributions are presented as a function of the angle with respect to the beam direction. The top and bottom panels correspond to the neutrino and anti-neutrino fluxes, respectively. The solid curves show the fluxes for the full angular range, and the dotted curves show the fluxes obtained after imposing an energy cut of $E_\nu>500~$GeV. For muon neutrinos and electron anti-neutrinos the main source is the beam muon decays(orange lines). 

The dominant contribution to tau (anti-)neutrinos comes from tau production in photon conversion in showers (dark blue lines), while electron and muon (anti-) neutrinos are predominantly produced via the $D$-meson production in photo-nuclear interactions in showers (light blue). We note that the contributions from $\mu^+\mu^-$ collisions (light and dark green lines) are approximately three orders of magnitude smaller. It is interesting to note that after the $E_\nu>500~$GeV cut, the dominant source of electron neutrinos and muon anti-neutrinos becomes $D$-meson production in neutrino interactions $\nu s\to\ell c$ (dotted light-purple curves in~\cref{fig:FluxesAngle}). This is because the electrons which result from shower neutrinos lose energy due to synchrotron radiation, thus generating neutrinos that are less forward and energetic, while neutrinos from the interaction of the neutrino beam produce very collimated and energetic secondary neutrinos.

\begin{figure}[t]
  \centering
    \includegraphics[width=1\textwidth]{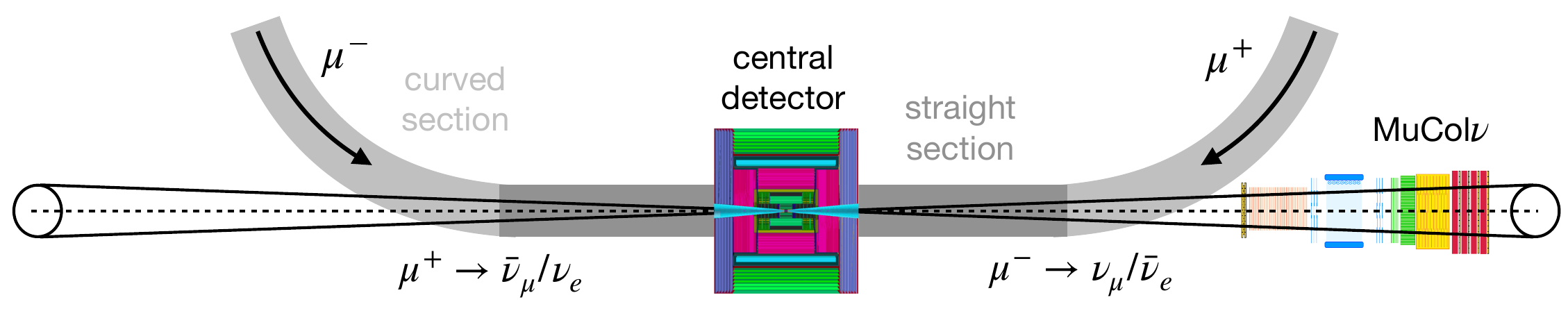}
  \caption{Schematic of a forward neutrino detector, MuCol$\nu$, placed $500$~m downstream of the interaction point. }
  \label{fig:location}
\end{figure}

From the angular distributions in~\cref{fig:FluxesAngle}, we see that the angular spread of the neutrino beam originating from muon decay is about $0.1$~mrad. Due to this strong collimation, even a relatively compact detector placed a few hundred meters downstream from the IP and centered around the beam axis will have excellent geometric acceptance. A possible setup is shown in~\cref{fig:location}: we consider a cylindrical target with a $25$~cm radius positioned at $z = 500$~m downstream from the IP, corresponding to an angular acceptance of $0.5~$mrad and a geometric acceptance of more than $80\%$ for neutrinos from muon decay. Detectors located farther away, but with proportionally larger transverse area, would be traversed by the same number of neutrinos. The distance of $500$~m was chosen to ensure at least $10$~m transverse separation from the muon beam in order to have sufficient space for both the experiments and shielding to sufficiently reduce beam induced backgrounds. 

\begin{figure}
    \centering
    \includegraphics[width=1\textwidth]{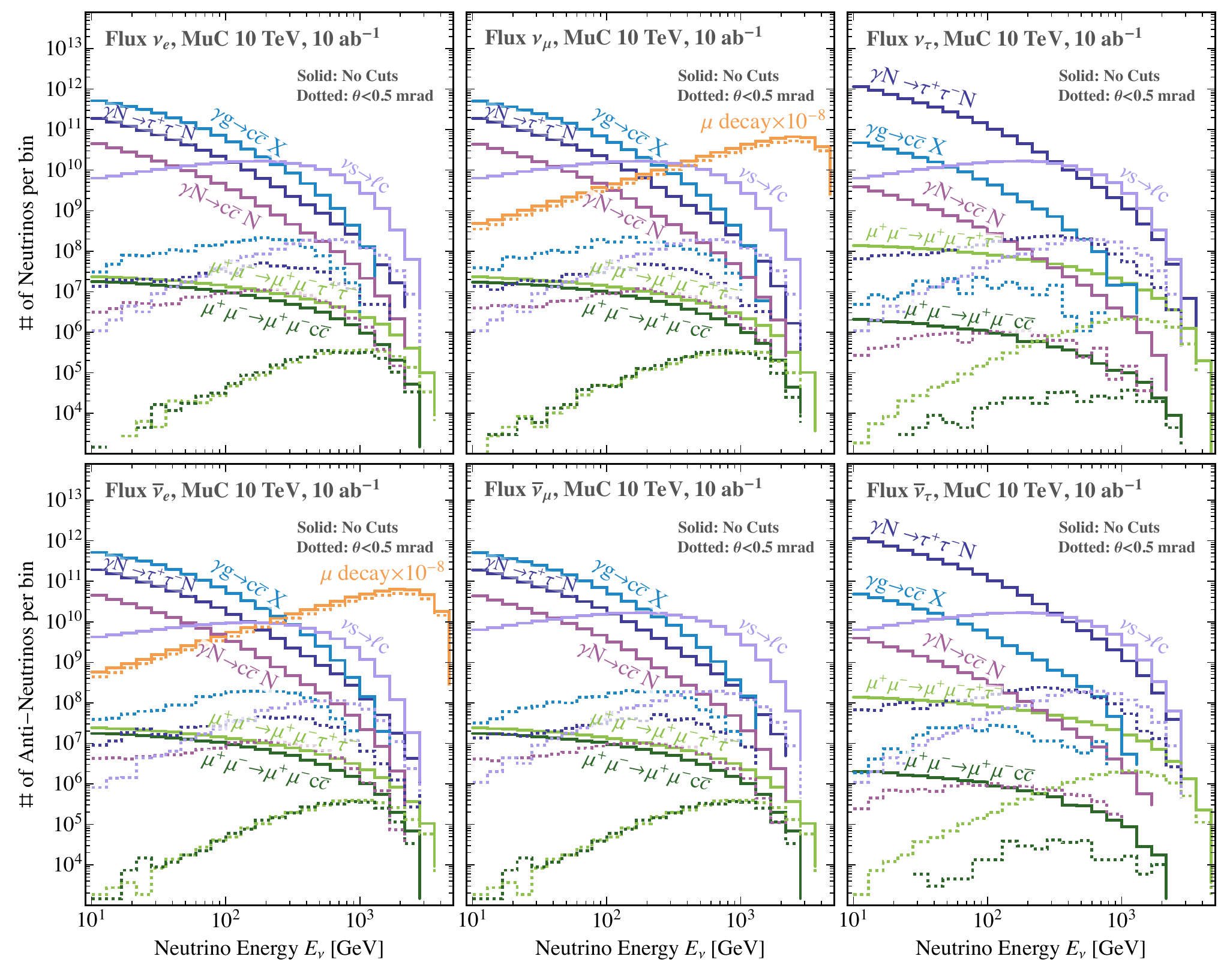}
    \caption{
    The fluxes for neutrinos (upper panels) and anti-neutrinos (lower panels)  produced in the straight section as a function of the neutrino energy, for different contributions from muon beam decay, showers and $\mu^+\mu^-$ collisions. The solid (dotted) curves show the fluxes before (after) the angular cut for the detector acceptance. The contribution from muon decay (orange) is multiplied by $10^{-8}$ to fit in the same plot.  The presented numbers assume a 10~year operation, corresponding to a $10~$ab$^{-1}$ of luminosity, at a \SI{10}{TeV} muon collider. 
    }
    \label{fig:FluxesEnergy}
\end{figure}

In~\cref{fig:FluxesEnergy}, we show the (anti-)neutrino fluxes as a function of the neutrino energy, where the solid curves correspond to the fluxes at the source, while the dotted lines denote to the energy spectra of neutrinos reaching the forward detector, after imposing an angular cut of $0.5~$mrad. We see that neutrinos from beam muon decay are so collimated that more than $80\%$ of them reach the forward detector after the angular cut. However, only less than $1\%$ of neutrinos from the showers or $\mu^+\mu^-$ collisions would reach such a detector; it is mostly low-energy neutrinos that fall outside the detector acceptance. Overall, most of the neutrinos originating from muon decay and interacting at the detector have energies above $1~$TeV, while neutrinos from other production modes are less energetic. 

At the energies of interest, neutrinos interact with the detector material through charged-current (CC) deep inelastic (DIS) terial through charged-current (CC) deep inelastic (DIS) neutrino-nucleus interactions, which neutrino-nucleus interactions, whose cross section rises linearly with the neutrino energy $E_\nu$ and can be approximately written as $\sigma \sim 10^{-35}~\text{cm}^2 \times E_\nu/\text{TeV}$. The corresponding anti-neutrino cross section is about a factor of two smaller due to the suppressed scattering on valence quarks. The probability of a neutrino interacting in the detector is then given by
\begin{equation}
    P = \frac{\sigma \times \text{number of nuclei}}{\text{detector area}} = \frac{\sigma}{ A_\text{det}}\ \frac{m_\text{det}}{m_n} \approx 3\times 10^{-12} \times \left(\frac{m_\text{det}}{\text{kg}}\right) \times  \left(\frac{E_\nu}{\text{TeV}}\right) \times  \left(\frac{2000~\text{cm}^2}{A_\text{det}}\right) ,
\end{equation}
where $m_\text{det}$ is the detector mass, $m_n$ is the mass of a nucleon, and $A_\text{det}$ is the detector's cross sectional area. When presenting numerical results, we use the cross sections obtained by \texttt{GENIE~3}~\cite{Andreopoulos:2009rq} based on the Bodek$-$Yang model~\cite{Bodek:2002vp}, which agree with more recent cross section calculations for TeV neutrinos, \texttt{NNSFv}~\cite{Candido:2023utz} and \texttt{CKMT+PCAC-NT}~\cite{Jeong:2023hwe}, to within $\lesssim 6\%$ over the range of energies of interest~\cite{FASER:2024ykc}. Convoluting the obtained fluxes with this interaction probability, we expect to detect roughly $10^{10}$, $1$, $1$, and $10^{-2}$ interaction events per ton and year of neutrinos originating from muon decay, EM showers, neutrino interactions and collisions respectively.

\begin{figure}[t]
  \centering
    \includegraphics[width=1\textwidth]{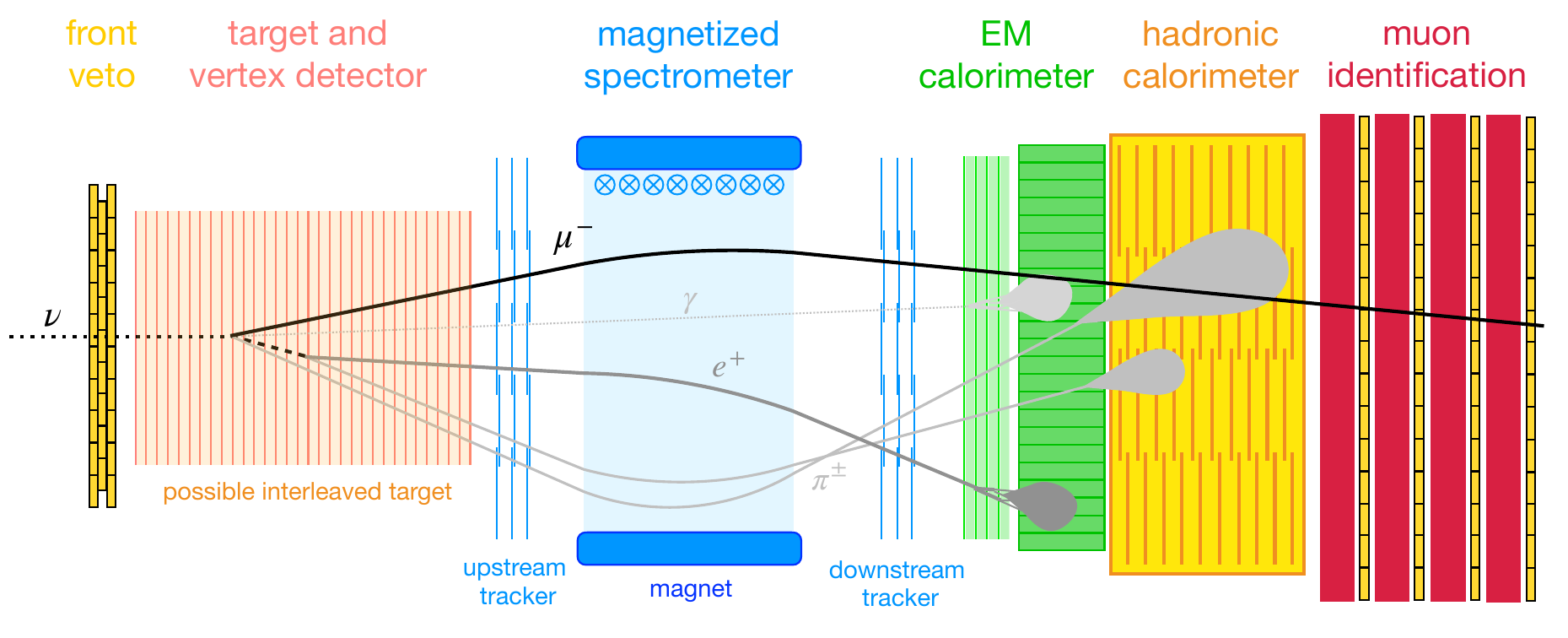}
  \caption{Conceptual layout of the MuCol$\nu$ detector. Located at the upstream end are a front veto station and a target consisting of multiple high-precision tracking layers, which also function as a vertex detector. Their spatial resolution enables reconstruction of the primary interaction vertex as well as displaced secondary vertices, providing sensitivity to short-lived particles such as charm hadrons and tau leptons. Downstream is a magnetized spectrometer based on the FASER2 design~\cite{Adhikary:2024nlv, Salin:2927003, FPF:2025bor}, consisting of a dipole magnet with tracking stations located upstream and downstream to accurately determine the charge and momentum of multi-TeV particles. The detector is completed by electromagnetic and hadronic calorimeters, followed by a dedicated muon identification system.}
  \label{fig:design}
\end{figure}

At present, no dedicated forward neutrino detector design exists for a muon collider. Nevertheless, a possible detector concept for muon storage rings and colliders was proposed in Ref.~\cite{King:1997dx}. The design features a $1$~m-long cylindrical vertex detector consisting of 750 silicon tracking layers, followed by a magnetized spectrometer, a calorimeter, and muon system. A similar design, adapting the design of the spectrometer and calorimeter systems of FASER2 at the Forward Physics Facility~\cite{Adhikary:2024nlv, Salin:2927003, FPF:2025bor}, was considered in Refs.~\cite{Adhikary:2024tvl, Kling:2025zsb} and referred to as MuCol$\nu$. A schematic design of this detector is illustrated in~\cref{fig:design}. This setup is designed to deliver precise momentum and energy measurements, together with efficient charged-particle identification. In particular, it enables the tagging of charm and bottom hadrons as well as $\tau$ leptons.

In this configuration, the silicon tracking layers also act as the neutrino interaction target, yielding a total target mass of approximately $100$~kg. The target mass, and therefore the neutrino interaction rate, could be significantly enhanced by inserting denser passive materials between the active tracking layers. However, the presence of additional material may degrade flavor-tagging performance, since hadronic scattering can produce displaced vertices that imitate the signatures of charm- or tau-particle decays. Furthermore, increasing the amount of passive material would raise the detector's optical depth from roughly 2.5 radiation lengths to considerably larger values. As a result, electrons would be largely absorbed before reaching the spectrometer, preventing charge determination and thereby eliminating the ability to distinguish between electron neutrinos and electron anti-neutrinos.

\begin{figure}
    \centering
    \includegraphics[width=1\textwidth]{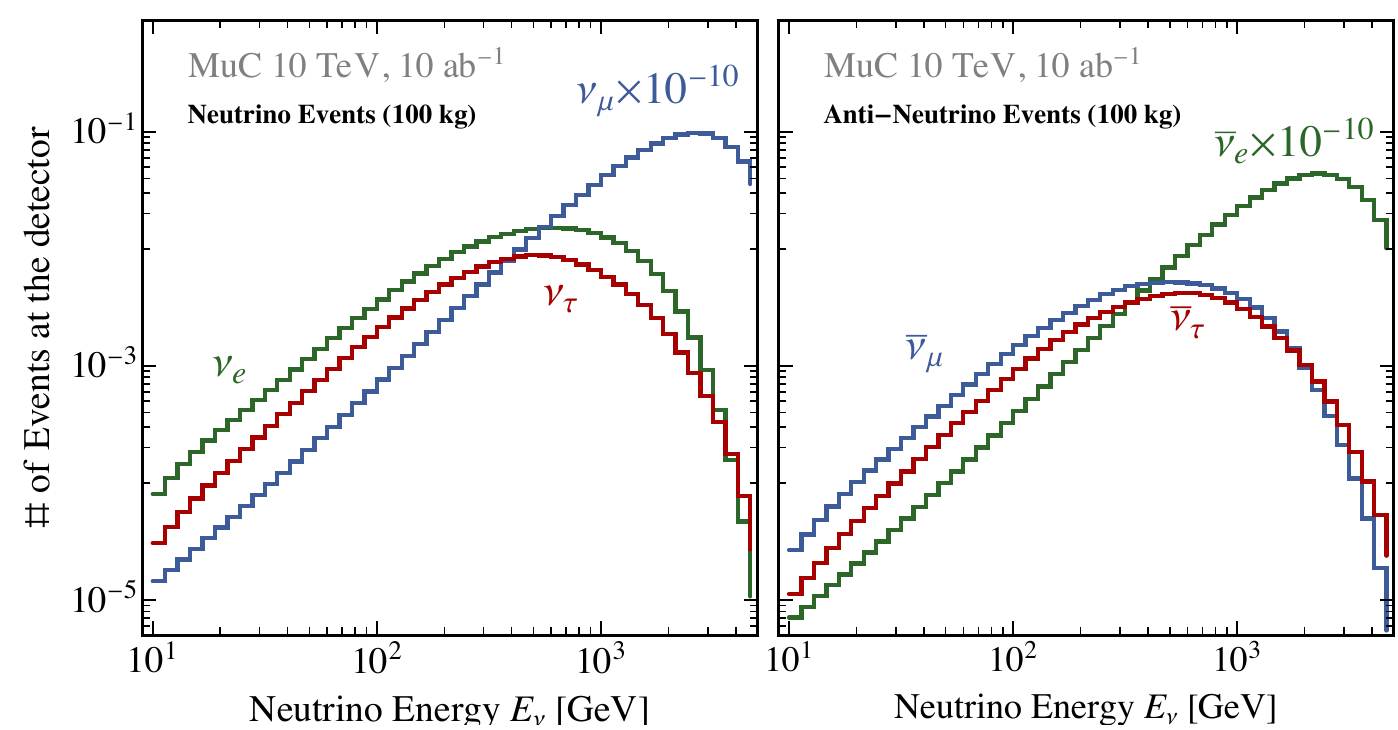}
    \caption{Predicted spectra of the  number of neutrino (left panel) and anti-neutrino (right panel) events in each bin as a function of  neutrino energy, interacting at the $100$~kg forward detector discussed in the text in the context of a \SI{10}{TeV} muon collider, taking data for ten years.
    }
    \label{fig:EventsEnergy}
\end{figure}

In \cref{fig:EventsEnergy} we show the energy spectra of neutrino (left panel) and anti-neutrinos (right panel) interactions in such a detector, assuming ten years of data taking, a target mass of \SI{100}{kg}, a cross sectional area of $\pi\times (\SI{25}{cm})^2$, and a location \SI{500}{m} downstream of the IP. We see that the majority of the neutrinos interacting in the forward detector have energies $E_{\nu} \gtrsim \SI{1}{TeV}$. In \cref{tab:numbers-summary} we summarize the expected interaction rates, breaking them down according to the different production channels and neutrino flavors. 
For $\nu_\mu$ and $\bar\nu_e$ the main contribution is from muon decay, as expected. For $\nu_e$ and $\bar\nu_{\mu}$ the electromagnetic showers and $D$-meson production from neutrino interactions contribute similarly. Electromagnetic showers dominate the production of $\nu_\tau$ and $\bar\nu_{\tau}$. It is interesting that, in spite of the high energies and short baselines, neutrino oscillations are not always negligibles. For $\nu_e$ and $\nu_\tau$, production via oscillations is at least an order of magnitude larger than the contribution from primary collisions.\\

\begin{table}[h!]
    \centering
   \begin{tabular}{c|c|c|c|c|c|c}
    \hline\hline
     &
     \multicolumn{6}{c}{Expected number of neutrinos at the detector} \\
    \hline
          & \quad\quad $\nu_e$
          &\quad $\bar\nu_e$
          &\quad $\nu_\mu$ 
          &\quad $\bar\nu_\mu$  
          &\quad $\nu_\tau$
          &\quad $\bar\nu_\tau$
           \\
    \hline
         $\mu$ decays  
         & - 
         & $5.2\cdot 10^{9}$ 
         & $1.1\cdot 10^{10}$ 
         & -
         & -
         & -
         \\
         EM showers  
         & $9.9\cdot 10^{-2}$
         & $4.9\cdot 10^{-2}$
         & $1.1\cdot 10^{-1}$
         & $4.7\cdot 10^{-2}$
         & $1.3\cdot 10^{-1}$
         & $6.2 \cdot 10^{-2}$
         \\
         $\nu$ interactions
         & $1.7\cdot 10^{-1}$ 
         & $4.6 \cdot 10^{-2}$ 
         & $1.7\cdot 10^{-1}$
         & $4.5\cdot 10^{-2}$
         & $1.0\cdot10^{-2}$
         & $7.2\cdot 10^{-3}$
         \\
         $\mu^+\mu^-$ collisions   
         & $8.6\cdot 10^{-4}$ 
         & $4.6\cdot 10^{-4}$
         & $8.0\cdot 10^{-4}$
         & $4.6\cdot 10^{-4}$
         & $3.3\cdot 10^{-3}$
         & $1.5 \cdot 10^{-3}$
         \\
         $\nu_\alpha\to\nu_\beta$ oscillation 
         & $1.9\cdot 10^{-3}$ 
         & -
         & -
         & $9.9\cdot 10^{-4}$
         & $2.5\cdot 10^{-2}$
         & $4.5\cdot 10^{-4}$
         \\
         \hline
         Total
         & $2.7\cdot 10^{-1}$ 
         & $5.2\cdot 10^{9}$
         & $1.1\cdot 10^{10}$
         & $9.3\cdot 10^{-2}$
         & $1.7\cdot 10^{-1}$
         & $7.1\cdot 10^{-2}$
         \\
    \hline\hline
\end{tabular}
\caption{Summary table of the number of expected CC neutrino interactions in a detector target of 100 kg with 10 years data taking.
}
\label{tab:numbers-summary}
\end{table}

\section{Physics Potential with High Neutrino Fluxes}
\label{sec:physics}

\begin{figure}
    \centering
    \includegraphics[width=1\textwidth]{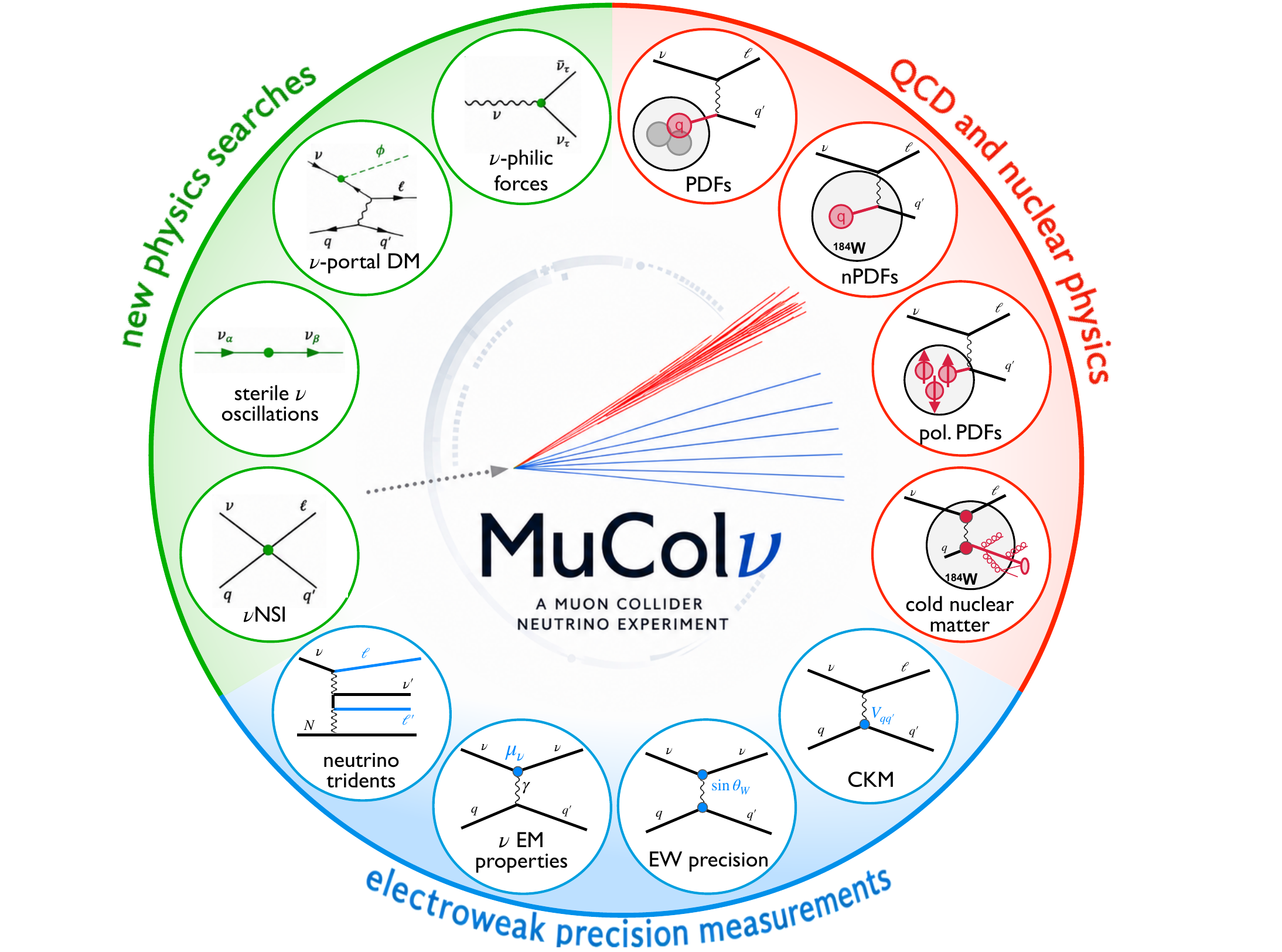}
    \caption{Examples for the physics opportunities afforded by a muon collider neutrino experiment (MuCol$\nu$) for precision measurements in QCD, electroweak physics, and new physics searches.}
    \label{fig:physics}
\end{figure}

The large expected neutrino event rate of about $10^{10}$ interactions per ton per year and the high purity of this sample provide a wide variety of opportunities for world-leading precision measurements in QCD and electroweak physics, as well as for searches for new physics. Several such opportunities are summarized in~\cref{fig:physics} and discussed below. \medskip

\noindent\textbf{QCD and Nuclear Physics:} Neutrino DIS data are a cornerstone of proton structure function determinations. The largest existing datasets were collected by NuTeV~\cite{NuTeV:2003kth}, CDHS~\cite{Berge:1989hr}, NOMAD~\cite{NOMAD:2003owt}, and CHORUS~\cite{CHORUS:2005cpn}, operating with neutrino beam energies of a few hundred GeV and recording about $10^6$ events each. MuCol$\nu$, with a $100$~kg target, would collect a comparable dataset within a few hours of nominal running while extending the kinematic reach towards lower $x$ and higher $Q$ by roughly an order of magnitude. Over $10$ years of operation, it would increase the available neutrino DIS statistics by four orders of magnitude, yielding a sample about two orders of magnitude larger than the electron DIS dataset accumulated at HERA.

Neutrino DIS can distinguish quark from anti-quark distributions through the charge of the final-state lepton and can provide direct sensitivity to the strange and charm PDFs via charm- and bottom-associated production~\cite{Faura:2020oom}. Measurements with targets spanning low-$Z$ to high-$Z$ nuclei would enable precise studies of nuclear effects, including shadowing, antishadowing, and the EMC effect. In this sense, MuCol$\nu$ would serve as a neutrino-ion collider, complementing the neutral-current program of the EIC~\cite{AbdulKhalek:2021gbh} with high-statistics charged-current measurements. The intense neutrino flux would also enable precision measurements with small polarized targets~\cite{Forte:2001ph}, such as the $1$~kg target used by COMPASS~\cite{COMPASS:2007rjf}, opening a new avenue for probing nucleon spin structure. Beyond nucleon structure, neutrino--nucleus collisions at MuCol$\nu$ would use nuclei as femtometer-scale detectors to investigate color neutralization and to systematically study the interplay of parton showers, hadronization, and cold nuclear matter effects~\cite{Accardi:2012qut}. These opportunities closely parallel and complement the broader EIC physics program~\cite{AbdulKhalek:2021gbh} and that of neutrino experiments at the LHC~\cite{Cruz-Martinez:2023sdv} and FCC~\cite{MammenAbraham:2024gun}. \medskip

\noindent\textbf{Electroweak Precision Measurements:} The large neutrino event sample at MuCol$\nu$ will enable precise measurements of DIS cross sections via the charged and neutral weak currents. Combined with the identification of final-state quark flavors, these measurements will allow precise determination of the CKM matrix elements. First studies indicate improvements in precision ranging from a factor of 3 for $V_{ub}$ to a factor of 44 for $V_{cd}$ relative to current determinations~\cite{Marzocca:2025inb}.

The high neutrino flux also enables precision measurements of electroweak observables, including the weak couplings $g_V$ and $g_A$ and the weak mixing angle $\sin^2\theta_W$. MuCol$\nu$ can measure $\sin^2\theta_W$ in several complementary channels, including neutrino DIS and elastic neutrino scattering on electrons and protons. Since these processes probe different momentum transfers and center-of-mass energies, their combination provides sensitivity to the running of the weak mixing angle. Ref.~\cite{deGouvea:2025zfq} shows that the expected sample of about $10^7$ elastic neutrino-electron scattering events would allow a measurement of $\sin^2\theta_W$ between momentum transfers of $10$~MeV and $1$~GeV with a relative precision of about $0.03\%$. The same data would also allow the electron and muon neutrino charge radii to be observed for the first time with high significance and measured with few-percent precision~\cite{deGouvea:2025zfq}, significantly surpassing the projected sensitivities of the DUNE near detector~\cite{Mathur:2021trm} and the LHC Forward Physics Facility~\cite{MammenAbraham:2023psg}.

Finally, the large rate of neutrino interactions would enable the study of rare processes such as neutrino trident production, $\nu N \to \nu \ell^+ \ell^- N$, which has not yet been conclusively observed. With about $10^6$ events expected, MuCol$\nu$ would not only observe this process for the first time, but would also be able to use it as a precision probe of electroweak parameters~\cite{Altmannshofer:2024hqd}. \medskip

\noindent\textbf{Searches for new physics:} The precise measurement of neutrino interaction cross sections at TeV energies provides a powerful probe of physics beyond the SM affecting neutrino interactions. In the presence of heavy new physics, its effects can be parameterized by four-fermion operators involving neutrinos, commonly referred to as non-standard interactions (NSI). Ref.~\cite{Kling:2025zsb} showed that MuCol$\nu$ can surpass the reach of current and planned low-energy precision experiments and the LHC for many such operators. With the 100-fold increase in neutrino flux found in this work, the projected bounds of Ref.~\cite{Kling:2025zsb} are expected to improve by about an order of magnitude for diagonal NSI and a factor of four for off-diagonal ones.

The large neutrino flux also makes MuCol$\nu$ well suited to search for sterile neutrinos with large mass-squared splittings through oscillations. With a baseline of $500$~m and neutrino energies around \SI{10}{TeV}, the experiment is sensitive to $\Delta m^2_{41}\sim10^3~\mathrm{eV}^2$, similar to LHC neutrino experiments~\cite{FASER:2019dxq,Bai:2020ukz}, but with substantially improved sensitivity due to the higher flux of the neutrino beam and reduced systematic uncertainties. In particular, appearance searches for otherwise absent neutrino flavors can probe mixing angles $\theta_{\alpha\beta}$ down to $\sim10^{-5}$, well beyond current constraints.

Similar appearance signatures also arise in other light new-physics scenarios. Ref.~\cite{Adhikary:2024tvl} studied a neutrino-philic mediator $\phi$ acting as a portal to dark matter. The study showed that MuCol$\nu$ can probe this scenario through the process $\nu_\mu N\to\mu^+\phi NX$ and improve existing constraints in the MeV--10~GeV mass range by up to two orders of magnitude. Another example is the decay of light feebly interacting particles into neutrinos, particularly the poorly constrained tau neutrino, as previously studied for the LHC~\cite{Kling:2020iar} and short-baseline neutrino experiments~\cite{Dev:2023rqb}.

\section{Summary and Conclusions}
\label{sec:sum}

In this paper, we have presented the first comprehensive calculation of high-energy neutrino fluxes expected at a \SI{10}{TeV} muon collider, using the most recent design of the interaction region. In addition to the main contribution coming from muon decays, we have also estimated neutrino production in the primary $\mu^+\mu^-$ collisions, in electromagnetic showers initiated by electrons from muon decay, and in secondary interactions of the beam neutrinos. Together, these processes generate neutrinos and anti-neutrinos of all three flavors and motivate a dedicated forward neutrino program at a muon collider.

After an introduction to the topics and the formulation of the neutrino production mechanisms, in Section {\ref{sec:collisions} we first presented the neutrino production in collisions of $\mm$ in a \SI{10} TeV muon collider. We found a substantial production rate in particular for the $\nu_\tau$ from photon-induced production in  the $\gg\to \ts, c\bar c$ channels. As shown in~\cref{fig:mupmumxsec}, we see that one may expect to reach 10 billion events with 10 ab$^{-1}$ luminosity. We also show many other contributing production channels in the figure, which are mostly subleading in contribution. Relevant to the neutrino flux in the forward region, we explore the $\ts$ kinematics, as shown in~\cref{fig:kindists}. Characteristic features include that the events are produced mostly near the kinematic threshold $\sim 2m_\tau$, and that the particle energies can extend to the order of \SI{10}{TeV}, and are quite forward with $\theta_{\rm beam}\sim 10^{-3}$ radians. 

The continuous decay of the beam muons produces by far the most intense neutrino beam. We re-evaluated this neutrino flux in \cref{sec:beam} using the latest design of a \SI{10}{TeV} muon collider \cite{MuCoL:2025quu,InternationalMuonCollider:2025sys}. Our simulation of the full interaction region gives approximately $\mathcal{O}(10^{19})$ muon neutrinos and electron anti-neutrinos per year in the straight section. This predicted flux is roughly $120$ times larger than that presented in Refs.~\cite{InternationalMuonCollider:2024jyv, InternationalMuonCollider:2025sys}. This enhancement reflects the approximately $500~\mathrm{m}$ length of the straight section and the increasingly collimated muon trajectories away from the interaction point. As the design of the interaction region is still evolving, the absolute flux remains subject to an uncertainty in the machine-design $\mathcal{O}(1)$. 

A second novel finding in this work is the neutrino flux generated by Michel-electron showers in the collider shielding, as described in \cref{sec:shower}. Tau pair production through photon conversion provides the leading shower induced source of $\nu_\tau$ and $\bar{\nu}_\tau$, while charm production in photon conversion and photo-nuclear interactions generates mainly $\nu_e$ and $\nu_\mu$ and their anti-neutrinos. The prompt decay of taus and charmed hadrons is essential: although muons and light hadrons are produced more copiously, their long lifetimes strongly suppress their neutrino yield in the collider environment. The electron-induced showers produce approximately $10^{12}$ neutrinos per year, while charm production in neutrino interactions in the surrounding rock (described in \cref{sec:secondary-nu}) adds a further contribution of order $10^{11}$.

In Section \ref{sec:flux}, we put our  results in perspective for a forward dedicated detector. In \cref{fig:FluxesAngle,fig:FluxesEnergy} we summarized the results of neutrino fluxes produced both at the interaction point and after the geometric cut reaching the detector. These results motivate a compact forward detector. We consider a muon collider neutrino detector, MuCol$\nu$, located at $z=500~\mathrm{m}$ and with a radius of $25~\mathrm{cm}$}. For a 100 kg target, the dominant beam flux gives approximately $10^{9}$ charged-current interactions per year. Over $10~\mathrm{ab}^{-1}$, the subleading sources yield several interactions of every neutrino flavor, including approximately $2.5$ tau-neutrino and tau-antineutrino events per ton of detector material.

A forward neutrino experiment would thus add a distinct intensity-frontier program to the energy-frontier mission of a muon collider.  In Section \ref{sec:physics} we discussed the physics opportunities as opened up by 1 billion neutrino interactions that can be recorded in a 100~kg detector per year. We argued that this enormous rate of neutrinos would enable a broad program of QCD and nuclear physics and would offer world leading constraints on nucleon and nuclear structure function determinations. It could significantly improve the precise determination of the CKM matrix elements and enable precision measurement of electroweak observables. It would also produce unprecedented samples of rare processes such as neutrino trident production. At the same time, the large flux and high beam purity would provide a powerful probe for new physics searches and enhance the sensitivity to non-standard neutrino interactions, sterile-neutrino oscillations, neutrino-philic forces, dark-sector portals, and other new particles. Our results therefore demonstrate that a dedicated forward detector would transform the unavoidable neutrino radiation of a muon collider into a unique experimental resource. \\ 

\noindent 
\textbf{Note Added: }{In the final stages of our manuscript preparation, we learned about independent work by~\cite{Hostert:2026}. We explore different aspects of forward muon flux at muon colliders. Our results are compatible and complementary.}

\section*{Acknowledgments}
We thank Matheus Hostert, Yang Ma, Toni Mäkelä, Federico Meloni, Laurie Nevay, Juan Rojo, and Keping Xie for useful discussions. The works of FB, TH and ZT were supported in part by the US Department of Energy under grant 
No.~DE-SC0007914 and in part by Pitt PACC. 
The research of WK is supported by the Deutsche Forschungsgemeinschaft (DFG, German Research Foundation) under grant 396021762 - TRR 257 and by Germany’s Excellence Strategy – Cluster of Excellence
“Color meets Flavor”, EXC 3107 – Project-ID 533766364. 
WK would also like to express his gratitude to Pitt PACC for support and to the people at Pitt PACC for their hospitality. 
JK would like to acknowledge support from the Cluster of Excellence ``Precision Physics, Fundamental Interactions, and Structure of Matter'' (PRISMA++ EXC 2118/2) funded by the German Research Foundation (DFG) within the German Excellence Strategy (Project ID 390831469). 
TH would like to thank the Aspen Center for Physics for hospitality during the final stage of this project, which is supported by NSF Grant PHY-2210452.

\FloatBarrier
\begin{appendices}
\appendix
\section{Flux comparisons for different lattice designs}
\label{App:Flux_Comparisions}

\begin{figure} [h!]
    \centering
    \includegraphics[width=1\textwidth]{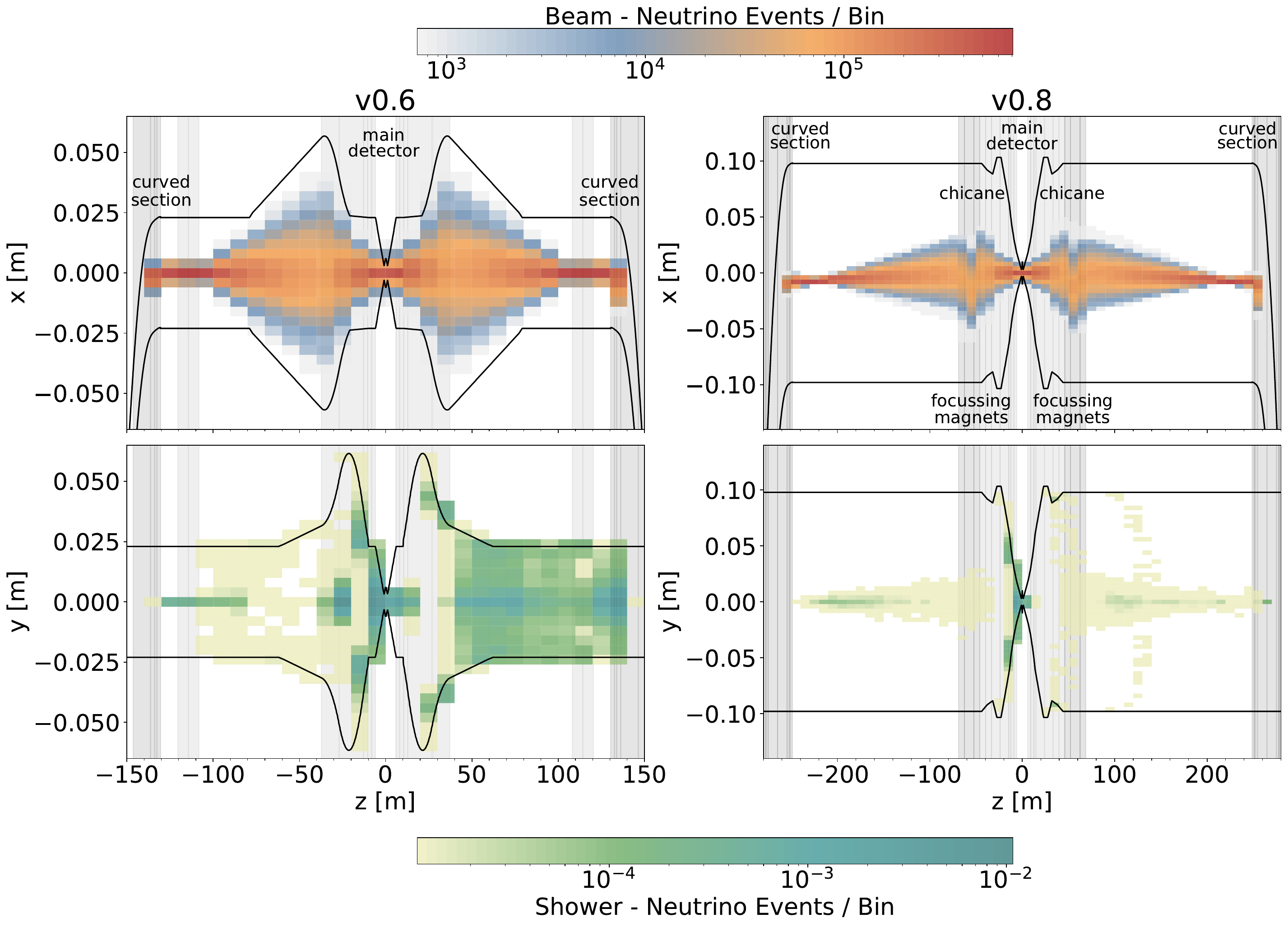}
    \caption{The boundaries of the beam pipe and magnets are shown in the horizontal (top) and vertical (bottom) planes, for interaction region designs v0.6 (left panels) and v0.8 (right panels). Similar to~\cref{fig:trajectory}, the shaded areas correspond to the quadrupole (light gray) and dipole (dark gray) magnets. Shown in a color is the distribution of production location for neutrinos interacting in the forward detector that were produced in muon decays (top) and electromagnetic showers (bottom).
    }
    \label{fig:beampipeCompare}
\end{figure}

The muon collider is still in the early stages of planning, and its design is therefore still being optimized. Throughout this paper, we have used design version~0.8 (v0.8)~\cite{MuCoL:2025quu}, as presented in the input to the 2025 European Strategy Update~\cite{InternationalMuonCollider:2025sys}, as our baseline. We note, however, that earlier designs differed significantly from this baseline and that further optimization is ongoing, likely resulting in additional design changes. Such modifications will naturally also affect the neutrino fluxes.

To quantitatively study the impact of design changes, we compare the neutrino fluxes obtained for the baseline design with those for the earlier version~0.6 (v0.6) presented in Ref.~\cite{Skoufaris:2023jnu} (with the v0.6 configuration files of Ref.~\cite{acc-models}). The corresponding interaction region for v0.6 is shown in the left panels of \cref{fig:beampipeCompare}. Located at $x=y=z=0$ is the primary IP, which is surrounded by the main detector, which extends to $|z|<6$. Downstream of the IP, quadrupole magnets located between $z=6 - 36$~m and $z=108 - 120$~m act as beam-focusing lenses. Together, they produce a beam angular divergence of 0.6~mrad at the IP while ensuring nearly parallel beam trajectories at the end of the straight section at $z=130$~m. The inner aperture of the beam pipe is shown as black line in~\cref{fig:trajectory}. As for v0.8, we use the design of the nozzle presented in Ref.~\cite{MAIA:2025hzm}. For the downstream part, less information is available in the literature. Here we follow the information provided in Ref.~\cite{Skoufaris:2022wwg, MuCoL:2024oxj, MuCoL:2025quu}, which assumes an inner radius of 2.3~cm in the curved part of the storage ring as well as a minimal aperture corresponding to five times the transverse spread of the beam. The latter was obtained from a simulated set of beam trajectories. 

The large number of muon decay into electrons along the straight section was identified as a source of radiation and background for the main detector. In order to suppress these backgrounds, a dipolar chicane was added before the final focusing quadrupole magnets in the v0.8 design, as illustrated in the right panels of~\cref{fig:trajectory}. The primary purpose of the chicane is to deflect electrons and reduce their energy due to synchrotron radiation, preventing them from reaching the nozzles. Apart from the addition of chicanes, the main difference between the two designs is the length of the straight sections. In the current design, the straight section extends to $|z| < \SI{230}{m}$, compared with $|z| < \SI{130}{m}$ in v0.6. Consequently, approximately twice as many muon decays are expected to occur within the interaction region, leading to a corresponding increase in the neutrino flux from muon decays in the current design.

\begin{figure}
    \centering
    \includegraphics[width=1\textwidth]{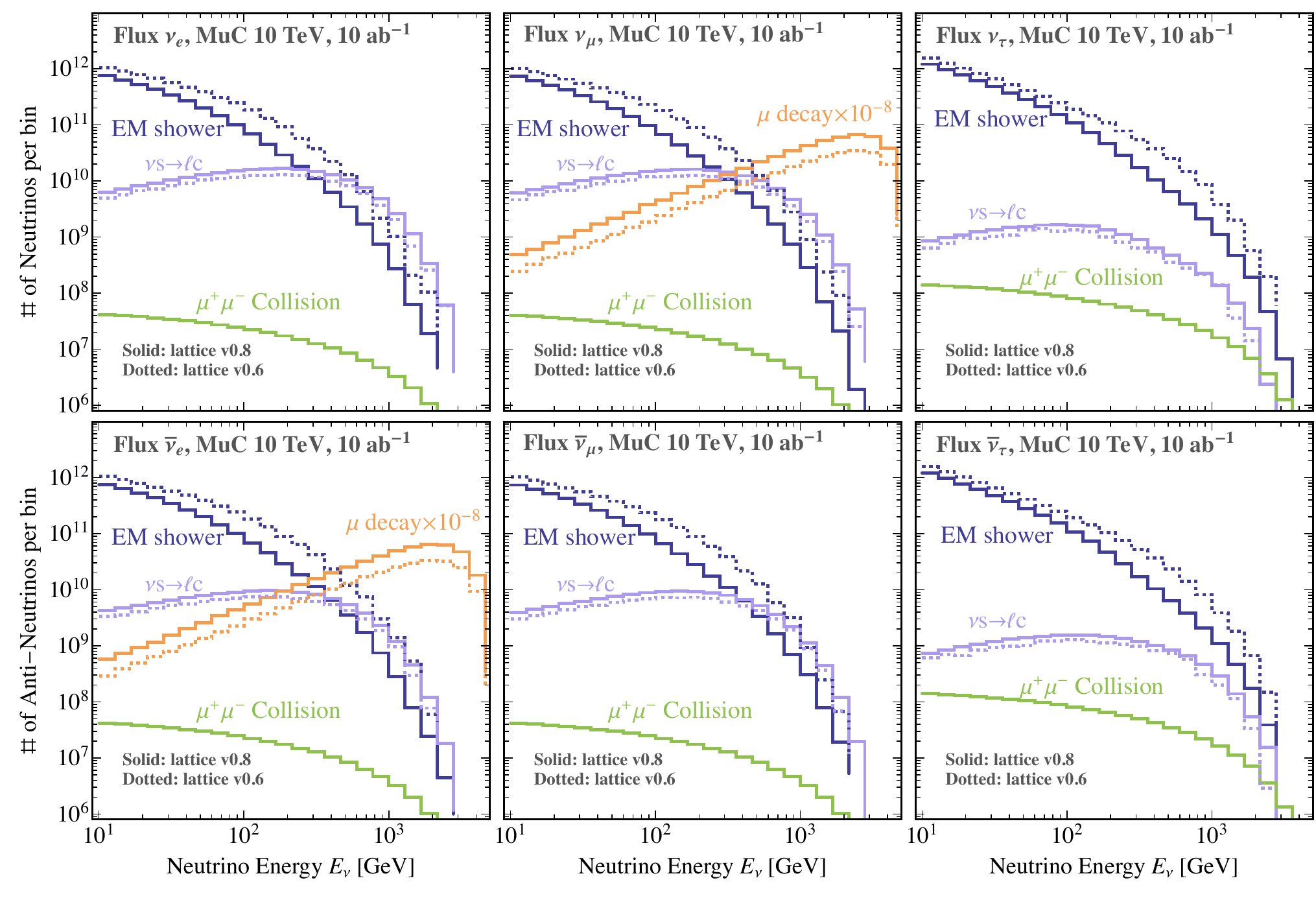}
    \caption{The fluxes for neutrinos (upper panels) and anti-neutrinos (lower panels) as a function of the neutrino energy for lattice designs v0.8 (solid curves) and v0.6 (dotted curves).
    }
    \label{fig:FluxesEnergyCompare}
\end{figure}

We compare the production location of muon decay neutrinos interaction at MuCol$\nu$ for the two lattice versions in the top panels of \cref{fig:beampipeCompare}. We see that in both versions the density of neutrino production in muon decays $dN/dV$ (with the volume $V$) is highest very close to the IP, as well as at the end of the straight sections close to the focusing quadrupole magnets right next to the beginning of the curved section.
Despite the assumed increase of the inner aperture of the beam pipe in v0.8 there is no apparent effect on the neutrino beam dynamics. Hence, the dominant effect between the two designs on the beam neutrinos comes from the length of the straight sections, and we expect to have twice more neutrinos using the current interaction region design. This is shown in orange in \cref{fig:FluxesEnergyCompare}, where we have shown the neutrino fluxes as a function of energy for the lattice v0.8 (v0.6) in solid (dotted) curves. With expecting more neutrinos from muon decay in v0.8, we can expect to have more D-meson production from neutrino interactions as well (see purple curves in \cref{fig:FluxesEnergyCompare}).

In contrast to neutrinos from muon decay, the addition of chicanes to the current lattice design significantly affects neutrino production from EM showers. It is expected, as the main motivation for adding the chicanes is deflecting electrons and not allowing them to reach the main detector. Hence, in the current design, mainly the $-20<z<0$ contribute to the EM production of neutrinos. This can be seen in the bottom right panel of \cref{fig:beampipeCompare}. As the bottom left panel show, the whole $|z| < \SI{140}{m}$ in v0.6 contributes in the EM neutrino production. Overall, this translates to approximately five times fewer EM shower neutrinos in v0.8 (see dark blue curves in \cref{fig:FluxesEnergyCompare}).

\FloatBarrier

\end{appendices}

\FloatBarrier
\bibliographystyle{JHEP}
\bibliography{references}

\end{document}